\documentclass[a4paper,10pt,DIV=12,flalign, abstracton]{scrartcl}

\setkomafont{disposition}{\fontfamily{bfu}}
\setkomafont{title}{\fontfamily{bfu}}

\usepackage[utf8]{inputenc}
\usepackage[T1]{fontenc}
\usepackage{amssymb,amsmath,amsthm}
\usepackage{nicefrac}
\usepackage{graphicx}
\usepackage{hyperref}
\usepackage{csquotes}
\usepackage{xcolor}
\usepackage[percent]{overpic}	
\usepackage{subfigure}
\usepackage{algorithm}				
\usepackage[noend]{algorithmic}			
\usepackage{siunitx}

 \usepackage[sort&compress,numbers]{natbib}

\makeatletter
\newcounter{ALC@tempcntr}

\newcommand{\LCOMMENT}[1]{%
	\setcounter{ALC@tempcntr}{\arabic{ALC@rem}}
	\setcounter{ALC@rem}{1}
	\item \textcolor{darkgray}{\hspace{-0.7cm} \emph{#1}:}
	\setcounter{ALC@rem}{\arabic{ALC@tempcntr}}
}%
\def\BState{\State\hskip-\ALG@thistlm}

\makeatother

\newcommand{\Moment}[2]{\left.{#1}\right|_{#2}}
\newcommand{\Hoehe}{H}		

\newcommand{\makro}[1]{\bar{#1}}
\newcommand{\mikro}[1]{#1}
\newcommand{\lRVE}{l_\mathrm{RVE}}
\newcommand{\proFak}{\lambda}							
\newcommand{\init}{^*}									
\newcommand{\plasMulti}{\kappa}							
\newcommand{\abs}[1]{\left|#1\right|}					
\newcommand{\Matnorm}[1]{\left\lVert#1\right\rVert}		
\newcommand{\Fnorm}[1]{\Matnorm{#1}}	
\newcommand{\Fliessfunk}{\Phi}							
\newcommand{\FliessSpa}{\sigma_\mathrm{Y}}				
\newcommand{\makroFliessSpa}{\makro{\sigma}_\mathrm{Y}}				
\newcommand{\Misesstress}{\sigma_{\mathrm{eq}}}
\newcommand{\dilatflowangle}{\makro\alpha_{\makro{n}}}
\newcommand{\Dissi}{D}							
\newcommand{\Aenderung}[1]{\dot{#1}}
\newcommand{\innerEner}{\Theta}

\newcommand{\Damage}{d}			
\newcommand{\DamageEner}{G_\Damage}

\newcommand{\NNfunc}{\boldsymbol{\mathrm{NN}}}

\newcommand{\NNfliess}{\NNfunc^{\makro{\Fliessfunk}}}
\newcommand{\NNdirec}{\NNfunc^{\makro{\breve{n}}}}
\newcommand{\Datenmenge}{\Omega}							
\newcommand{\trainData}{{\Datenmenge_\mathrm{T}}}				
\newcommand{\valiData}{{\Datenmenge_\mathrm{V}}}				

\newcommand{\Bias}{B}
\newcommand{\Weight}{W}

\newcommand{\AnzahlNeuron}[1]{{N_\mathrm{#1}}}
\newcommand{\inpNeuron}{\AnzahlNeuron{i}}
\newcommand{\hidNeuron}{\AnzahlNeuron{h}}
\newcommand{\outNeuron}{\AnzahlNeuron{o}}

\newcommand{\rindent}{r_{\mathrm{t}}}
\newcommand{\uindent}{u_{\mathrm{t}}}

\title{A hybrid approach to simulate the homogenized irreversible elastic-plastic deformations and damage of foams by neural networks}

\author{Christoph Settgast, Geralf Hütter, Meinhard Kuna, Martin Abendroth}

\begin{document}

\maketitle

\begin{abstract}

Classically, the constitutive behavior of materials is described either phenomenologically, or by homogenization approaches. Phenomenological approaches are computationally very efficient, but are limited for complex non-linear and irreversible mechanisms. Such complex mechanisms can be described well by computational homogenization, but respective FE$^2$ computations are very expensive.

As an alternative way, neural networks have been proposed for constitutive modeling, using either experiments or computational homogenization results for training. However, the application of this method to irreversible material behavior is not trivial.

The present contribution presents a hybrid methodology to embed neural networks into the established framework of rate-independent plasticity. Both, the yield function and the evolution equations of internal state variables are represented by neural networks. Respective training data for a foam material are generated from RVE-simulations {under monotonic loading}.
It is demonstrated that this hybrid multi-scale neural network approach (HyMNNA) allows to simulate efficiently even the anisotropic elastic-plastic behavior of foam structures with coupled anisotropic evolution of damage and non-associated plastic flow.

\end{abstract}

\section*{Graphical abstract}
%
\begin{figure}[h!]
 	\centering
 	\includegraphics[width=\textwidth]{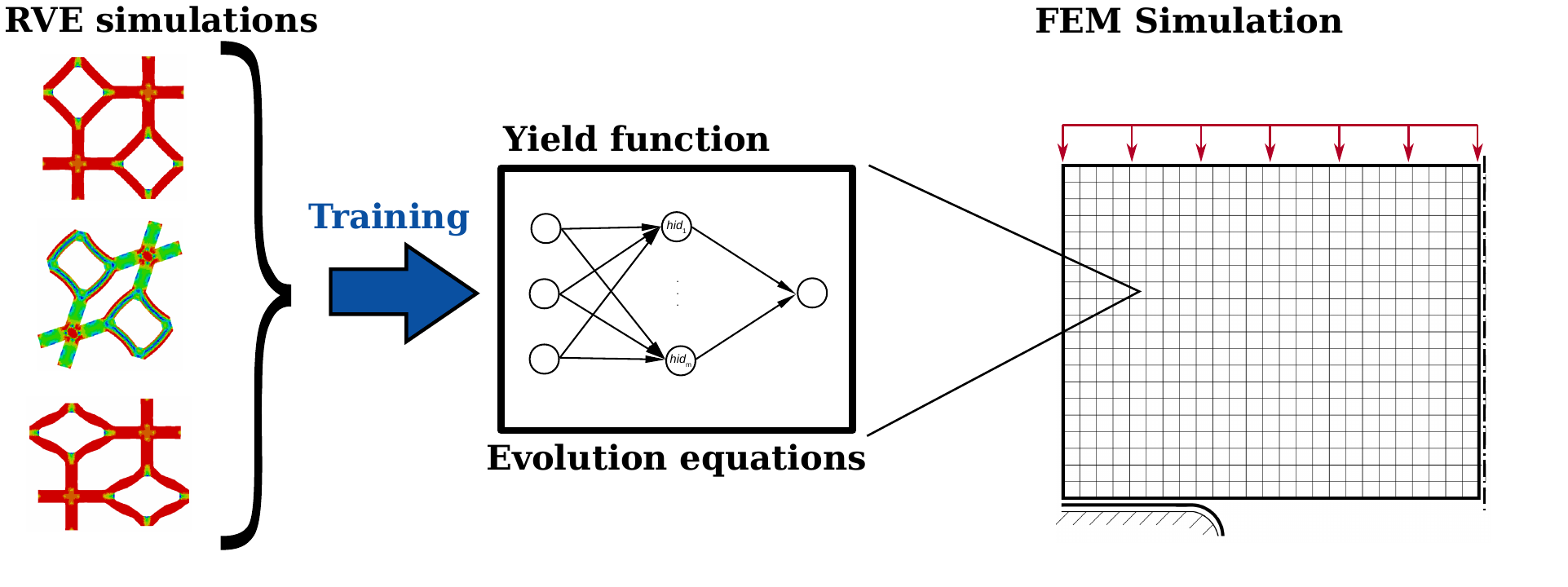}
\end{figure}

\section{Introduction}
\label{sec:introduction}

The formulation of constitutive relations for particular materials has been a challenging key task in solid mechanics modeling, in particular when nonlinear and irreversible behavior is considered. 
Traditionally, constitutive relations can be obtained on two ways. In the macroscopic phenomenological way, the general structure of the constitutive relations is postulated based on a number of \emph{qualitative} features, like isotropy or rate dependency. Subsequently, {a} \emph{finite} number of constitutive parameters is calibrated from respective experiments. An overview over phenomenological theories of plasticity can be found e.~g.\ in \citep{Chaboche2008}.

Alternatively, the macroscopic constitutive behavior can be obtained from the behavior of the microstructure by \emph{homogenization}, provided that sufficient information about the microstructure are available. 
For the homogenization, a boundary-value problem needs to be solved for the micro-structure at each macroscopic material point. For a linear-elastic material, the solution of the microscopic problem can be obtained simply and cheaply by superposition of a few 
{load cases. The latter can be computed \emph{offline}, i.~e.\ before the the macroscopic problem is being solved.} 
Unfortunately, such a superposition is not possible for nonlinear material behavior, but the microscopic problem has to be solved \emph{online} for each material point, {i.~e.\ simultaneously with the macroscopic problem}. The implementation of such an approach into the finite-element method is known as FE$^2$ \citep{Feyel1999,Miehe1999}. The FE$^2$ method is very universal and has been employed for many problems. 
The disadvantage of the FE$^2$ method is its huge computational expense. Although a number of approaches have been proposed to improve its computational efficiency, FE$^2$ simulations are still orders of magnitude more expensive than simulations with phenomenological constitutive relations. Reviews on homogenization and multi-scale methods were given by \citet{Kanoute2009} in context of composite materials and by \citet{Besson2010} with respect to ductile damage.

More than two decades ago, it was proposed to use neural networks as a third way for the representation of the constitutive relations. 
A review on neural networks and machine learning in continuum materials mechanics was given recently by \citet{Bock2019}.
This approach is appealing insofar as it should be able to represent complex material behavior without the long and usually iterative way of formulating constitutive relations and identifying the parameters therein. Rather, the constitutive response should result directly from the training data. The training itself may be computationally expensive, but it can be done \emph{offline}, whereas the required online computation of the response function is not more expensive than conventional phenomenological models.
Either experimental data can be used for training \citep{Ghaboussi1998ComputGeotech,Javadi2009AdvEngInform,Al-Haik2006,Zopf2017} or RVE simulations \citep{Le2015,Settgast2019,Fritzen2019,Lu2018,Liu2019,Hambli2011,Yang2018}.

Though, it turned out that the choice of the response function to be represented by neural networks within the constitutive framework is not trivial. Several authors \citep{Settgast2019,Fritzen2019,Shin2000,Zopf2017} trained the non-linear stress-strain relationship, corresponding to a pseudo-plastic Cauchy-elastic material behavior. 
In a similar way, a non-linear conductivity problem was addressed by neural networks \citep{Lu2018}. 
{\citet{Kirchdoerfer2016} minimized the distance between actual states (of stress and strain) and training data directly within their \enquote{data-driven computing} approach, assuming that the material possesses an inherent Cauchy-elastic relation.}
Instead, the strain-energy density of a non-linear hyper-elastic material was represented by neural networks in \citep{Le2015,Liang2008EngStruct}. A hypo-elastic rate formulation was represented by  \citet{Javadi2009AdvEngInform}.

The representation of the constitutive relations by neural networks becomes even more challenging when irreversible material behavior is considered. \citet{Ghaboussi1998ComputGeotech} proposed to represent the current stress or its rate as a trained function of the strain and its rates. 
Alternatively, \citet{Al-Haik2006} represented the non-linear relaxation function of a visco-elastic composite by neural networks.
Such nonlinear ODEs or convolution integrals in time, or discretized versions thereof, are very general representations of irreversible material behavior. They comprise any kinds of anisotropic and rate dependent behavior.
The problem with such a general representation is that a correspondingly large set of training data is required, covering loading in different directions at different rates with intermediate unloading etc. 
 
However, certain \emph{qualitative} features of the material behavior are already known {in many cases}, e.~g.\ spatial symmetries (isotropic, cubic, etc.), a tension-compression symmetry, rate-independent behavior or the presence of an elastic domain. Or it may happen that certain features are negligible for the envisaged application. Incorporating such known features a priori (and/or neglecting other ones) reduces the amount of required training data considerably. Vice verse, it may happen that the amount of training data is limited such that certain features need to be postulated to generate a constitutive relation for specific kinds of loading. For instance, tension-compression symmetry needs to be postulated if only training data in the compression regime are available. 

In this sense, {\citet{Shen2004} and} \citet{Liang2008EngStruct} trained the strain-energy function directly by the strain invariants to model the isotropic hyper-elastic behavior of {rubber and foam materials, respectively}. \citet{Jung2006ComputStruct} incorporated isotropy ad-hoc into their neural network representation of a visco-elastic material. 
{\citet{Zopf2017} employed the micro-sphere model to describe the multi-axial behavior of rubber, based solely on the uni-axial behavior of a single polymer chain. The latter was represented by a recurrent neural network, whose training requires only uni-axial tests.}
\citet{Furukawa2004EngAnalBoundElem} formulated the yield condition of a material ad-hoc to account for rate-independent behavior with an elastic domain, but trained the evolution of drag stress and back stress to represent the complex combined hardening under uni-axial cyclic loading. 
In a similar way, \citet{Li2019} employed a Hill yield condition and represented the dependency of the equivalent yield stress on temperature and strain-rate by a neural network.
{\citet{Ibanez2019} used a sparse identification technique to represent the yield surface itself. In \citep{Ibanez2018}, sampling points of the yield surface were employed directly within a data-driven approach.}
Though, in general the shape of the yield surface evolves itself with ongoing multi-axial deformations. 
This so-called distortional hardening is known to be relevant for foam materials \citep{Deshpande2000,Miller2000a,Storm2016IntJMechSci} or sheet metal forming, and its modeling is very challenging, see e.~g.\ \citep{Feigenbaum2007,Manopulo2015,Borgqvist2014,Pietryga2012} and references therein. Also non-associated plastic flow has been diagnosed for foams and sheet metals \citep{Zhang1998,Forest2005a,Li2019}.

The scope of the present contribution is to establish a hybrid approach between conventional phenomenological plasticity and neural network constitutive modeling whose basic idea has been outlined recently by the authors at the ECCM conference \citep{Settgast2018ECCM}.
The existence of a yield condition to separate elastic and plastic domain is specified ad hoc, as well as a set of internal state variables. However, the particular yield function and the evolution equations are \emph{not} specified ad hoc, but only the functional dependencies for a representation by neural networks. This methodology is applied to the {rate-independent} plastic and damage behavior of foam material {under monotonic loading within a small deformation framework}.

The present contribution is structured as follows. Section~\ref{sec:microstructure} presents the microscopic model of the foam material, before the macroscopic model is developed in section~\ref{sec:macromodel}. A special focus is put on the extraction of the required training data from computational homogenization. Subsequently, the hybrid multi-scale neural network approach (HyMNNA) is applied in section~\ref{sec:application} to simulate the material behavior at a single point as well as the inhomogeneous deformations of a foam structure. Finally, section~\ref{sec:summary} closes with a short summary and an outlook.

\section{Microstructure}
\label{sec:microstructure}

\begin{figure}
	\centering
	\includegraphics[width=0.8\textwidth]{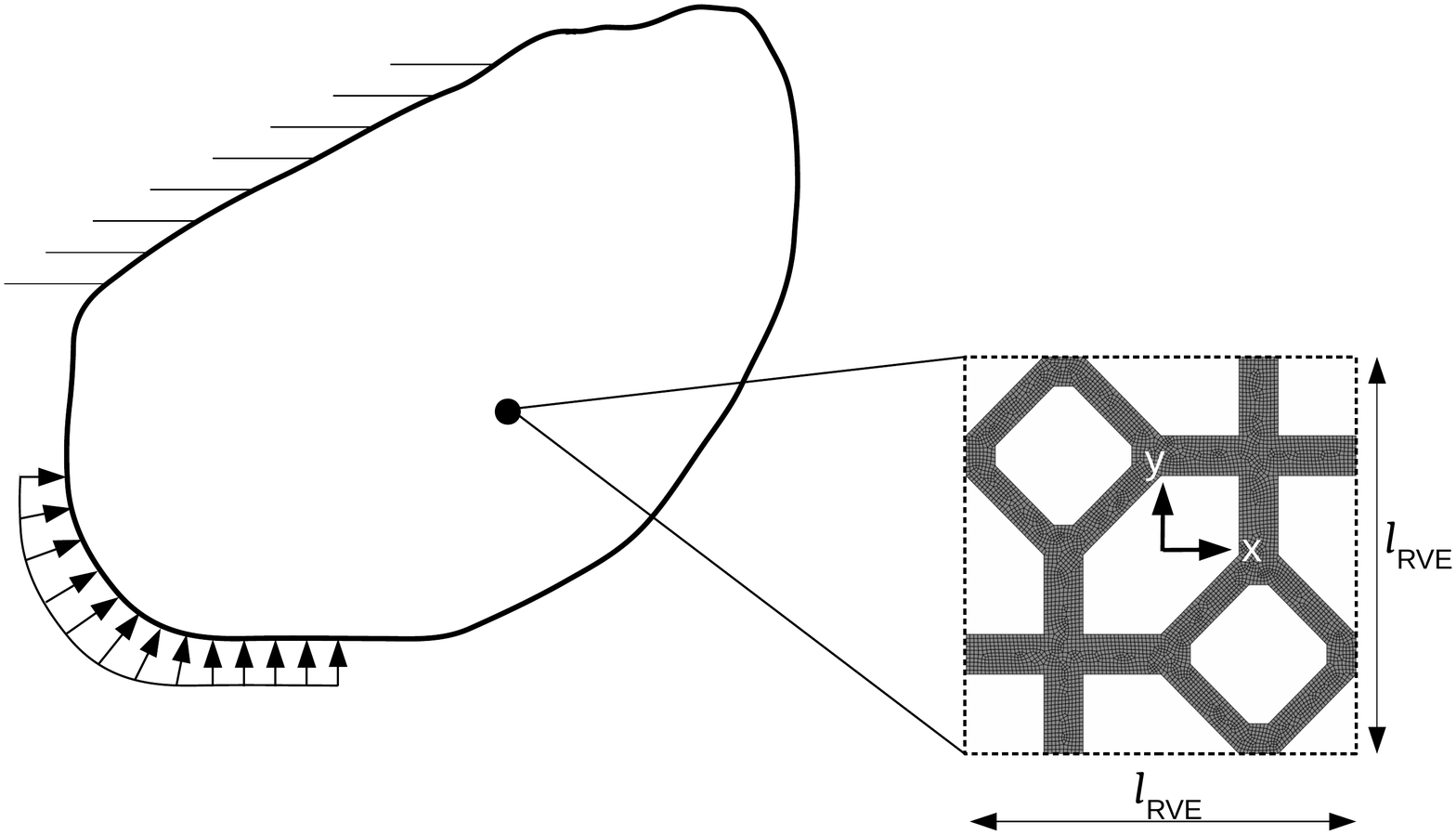}
	\caption{Homogenization with RVE of foam structure including employed FEM mesh}
	\label{fig:Theorie_Homogen_MikroMakro}
\end{figure}

For simplicity, an idealization of a foam material is considered for the present proof-of-concept study in form of a periodic arrangement of 2D projections of a Kelvin cell as shown in \figurename~\ref{fig:Theorie_Homogen_MikroMakro} within a small deformation framework. This cell has a relative density of 36.5\%, cf.~\citep{Settgast2019}.

\subsection{Bulk material}

The bulk material of the foam at the microscale is characterized by Hooke's law 
\begin{equation}
	\sigma_{ij} = C_{ijkl}^0\varepsilon_{kl}^\mathrm{el} = C_{ijkl}^0\left(\varepsilon_{kl}-\varepsilon_{kl}^\mathrm{pl}\right).
\label{equ:Hooke}
\end{equation}
between stresses $\sigma_{ij}$ and elastic strains $\varepsilon_{kl}^\mathrm{el}=\varepsilon_{kl}-\varepsilon_{kl}^\mathrm{pl}$. Isotropic behavior of the bulk material is assumed such that the stiffness tensor
\begin{equation}
	C_{ijkl}^0 = \frac{\mikro{E}}{1+\mikro\nu}\left[\frac{\mikro\nu}{1-2\mikro\nu}\delta_{ij}\delta_{kl}+\frac{1}{2}\left(\delta_{ik}\delta_{jl}+\delta_{il}\delta_{jk}\right)\right]
\end{equation}
is given uniquely in terms of Young's modulus $\mikro{E}$ and Poisson ratio $\nu$, respectively.
The Mises yield function
\begin{equation}
	\mikro{\Fliessfunk}\left(\mikro\sigma_{ij}\right) = \Misesstress - \mikro\FliessSpa
	\label{equ:Theorie_MatMod_Bsp_Fliess}
\end{equation}
is employed for the plastic behavior with an associated flow rule
\begin{equation}
	\Aenderung{\varepsilon}_{ij}^\mathrm{pl} = \plasMulti\frac{\partial \Fliessfunk}{\partial \sigma_{ij}}\,.
	\label{equ:Theorie_PlasDehn}
\end{equation}
Therein,
$\Misesstress$
refers to the Mises equivalent stress. Rate-independent plastic flow is implemented by the loading-unloading condition (\textsc{Karush-Kuhn-Tucker} condition)
\begin{equation}
\Fliessfunk\le 0, \quad \plasMulti\ge 0\quad\text{{and}}\quad \Fliessfunk\,\plasMulti=0\,,
\label{equ:load-unload}
\end{equation}
stating that the plastic multiplier $\plasMulti$ can take values different from zero only if the yield function attains a value of zero $\Fliessfunk=0$.
The microscopic dissipation amounts to
\begin{equation}
	\Aenderung{\mikro\Dissi}=\sigma_{ij} \Aenderung{\varepsilon}_{ij}^\mathrm{pl}\,.
	\label{equ:dissipationmicro}
\end{equation}
Ideal plasticity $\mikro\FliessSpa=\text{const.}$ is considered in the following  
under plane stress conditions. The yield stress of the bulk material amounts to  $\FliessSpa=0.061\,E$ and the Poisson ratio is chosen as $\nu=0.14$. The following results are presented in a normalized way that is invariant to the choice of $\FliessSpa$ and $E$.

\subsection{Damage}
\label{sec:damagemicro}

A second set of training data is created which involves damage at the microscale. For this purpose,
Hooke's law~\eqref{equ:Hooke} is extended towards isotropic damage
\begin{equation}
	\sigma_{ij}= (1-\mikro{\Damage})C_{ijkl}^0 \left(\varepsilon_{kl}-\varepsilon_{kl}^\mathrm{pl}\right)
	\label{equ:Theorie_SchaedigungCijkl}
\end{equation} 
of Lemaitre-Kachanov type. The scalar damage variable $\mikro{\Damage}$ is amended to the yield stress in the modified yield function
\begin{equation}
	\Fliessfunk = \Misesstress -(1-\mikro{\Damage})\FliessSpa
	\label{equ:Theorie_SchaedigungFliessSpa}
\end{equation}
as well. Evolution of damage is described by the established approach of \citet{Hillerborg1976CementConcreteRes}
\begin{equation}
	\dot{\mikro{\Damage}} = \begin{cases}
	 							\frac{\dot{\mikro\varepsilon}_\mathrm{eq}^\mathrm{pl}}{\varepsilon_{\mathrm{ref}}} & \text{if }\mikro\varepsilon_\mathrm{eq}^\mathrm{pl} \geq \varepsilon_\mathrm{eq}^{\mathrm{pl}_0} \\
	 				   		    0 & \text{else}
	 						\end{cases}
	 \label{equ:Theorie_Schaedigungsentw}
\end{equation}
Therein, $\dot{\mikro\varepsilon}_\mathrm{eq}^\mathrm{pl}=\sqrt{2/3\dot{\mikro\varepsilon}_{ij}^\mathrm{pl}\dot{\mikro\varepsilon}_{ij}^\mathrm{pl}}$ is the rate of equivalent plastic strain. Furthermore, $\varepsilon_\mathrm{eq}^{\mathrm{pl}_0}$ and $\varepsilon_{\mathrm{ref}}$ refer to a equivalent plastic strain for damage initiation and to a reference strain for the softening regime, respectively. The pathological mesh sensitivity of such a local damage model within an finite element implementation can be reduced by relating the reference strain to the size $l_\mathrm{Element}$ of the respective elements as $\varepsilon_{\mathrm{ref}}=2\mikro\DamageEner/(\mikro\FliessSpa l_\mathrm{Element})$.
In the following, the parameters are chosen as $l_\mathrm{Element}=0.01\,\lRVE$, $\varepsilon_\mathrm{eq}^{\mathrm{pl}_0}=0$ and $\DamageEner=1.5\,(\FliessSpa)^2\,\lRVE/E$, respectively.

\subsection{Computational homogenization}

In the established homogenization theory of Hill, macroscopic stress and strain are introduced as volume averages over their microscopic counterparts
\begin{align}
	\makro\sigma_{ij} &= \frac{1}{V_\mathrm{RVE}} \underset{ V_\mathrm{RVE}}{\int} \mikro\sigma_{ij} \mathrm{d}V\,, &
	\makro\varepsilon_{ij} &= \frac{1}{V_\mathrm{RVE}} \underset{ V_\mathrm{RVE}}{\int} \mikro\varepsilon_{ij} \mathrm{d}V
	\label{equ:Theorie_Homogen_makroDehnVolum}	
\end{align}
over the representative volume element $V_\mathrm{RVE}$, as shown in \figurename~\ref{fig:Theorie_Homogen_MikroMakro}.
Periodic boundary conditions
\begin{equation}
	u_i = \makro\varepsilon_{ij}x_j + \tilde{u}_i \qquad \text{at} \qquad \partial V_\mathrm{RVE}
	\label{equ:Theorie_Homogen_RB}
\end{equation}
with periodic fluctuations
\begin{equation}
	\tilde{u}_i(x_k^{-})=\tilde{u}_i(x_k^{+}).
	\label{equ:Theorie_Homogen_homolog}
\end{equation}
ensure that the macro-homogeneity condition (Hill-Mandel condition) is satisfied, {cf.\ e.~g.\ \citep{Feyel1999,Miehe1999}}. Therein, $x_k^{-}$ and $x_k^{+}$ refer to position vectors of of homologous points at the boundary of the RVE. 
Elimination of fluctuations $\tilde{u}_i$ from Eq.~\eqref{equ:Theorie_Homogen_homolog} using Eq.~\eqref{equ:Theorie_Homogen_RB} yields the relation
\begin{equation}
	u_i(x_j^+)-u_i(x_j^-)=\makro\varepsilon_{ij}\left[x_j^+ - x_j^-\right]\,.
	\label{equ:Theorie_periodRB}
\end{equation}
In the finite element model for the Kelvin foam shown as inset of \figurename~\ref{fig:Theorie_Homogen_MikroMakro}, Eq.~\eqref{equ:Theorie_periodRB} is implemented as multi-point constraint for each set of homologous nodes at the boundary \citep{Feyel1999}.
Furthermore, the macro-homogeneity condition implies that the macroscopic dissipation
\begin{equation}
	\Aenderung{\makro\Dissi}=\frac{1}{V_\mathrm{RVE}} \underset{ V_\mathrm{RVE}}{\int} \Aenderung{\mikro\Dissi} \mathrm{d}V
	\label{equ:dissipationmicromacro}
\end{equation}
as source in the macroscopic energy balance needs to be identified as volume average as well \citep{Chatzigeorgiou2016}.

\section{Macroscopic model using neural networks}
\label{sec:macromodel}

\subsection{Elastic-plastic behavior}
\label{sec:elasticplasticmacro}

The microscopic model presented in the previous section is rate-independent and each material point possesses an elastic domain within which the material behaves linear elastic. Thus, it is known that the macroscopic behavior \enquote{inherits} these properties. Consequently, the macroscopic law of state is constructed in the same way as
\begin{equation}
	\makro\sigma_{ij}  = \makro C_{ijkl}\,\makro\varepsilon_{kl}^\mathrm{el} = \makro C_{ijkl}\left(\makro\varepsilon_{kl} - \makro\varepsilon_{kl}^\mathrm{pl}\right)
	\label{equ:statelawmacro}
\end{equation}
in combination with a macroscopic yield function $\makro{\Fliessfunk}$ and a loading-unloading condition analogous to \eqref{equ:load-unload}. In particular, the yield function
\begin{equation}
	\makro{\Fliessfunk} = \Fnorm{\makro\sigma_{ij}}  - \NNfliess\left(\makro{I}_1,\makro\varepsilon_\mathrm{eq}^\mathrm{pl},\abs{\makro\varepsilon_{12}^\mathrm{pl}}\right),
	\label{equ:MakElasPlas_MatMod_hybrid_Fliessfunk}
\end{equation}
is constructed in terms of the Frobenius norm $\Fnorm{\makro\sigma_{ij}}=\sqrt{\makro\sigma_{ij}\makro\sigma_{ij}}$ of the macroscopic stress tensor. The reason for this choice is that values of the macroscopic yield function $\makro{\Fliessfunk}$ cannot be extracted from the RVE simulations. Rather, it can be detected only whether plastic yielding {is} active corresponding to $\makro{\Fliessfunk}=0$, or whether the current response is elastic $\makro{\Fliessfunk}<0$. That is why the Frobenius norm is favorable for the generation of suitable training data for the equivalent yield stress function $\NNfliess$ as will be explained in detail below in section~\ref{sec:training}.

The main part of constitutive modeling within the present hybrid approach is to formulate suitable input values for the neural network representation of the equivalent yield stress function $\NNfliess$.
Firstly, this function is assumed to depend on the first stress invariant $\makro{I}_1=\makro\sigma_{kk}$, since it is known to have a significant influence on the plastic behavior of foams. {The third invariant vanishes identically for the plane problem under consideration.} Secondly, the equivalent plastic strain $\makro{\varepsilon}_\mathrm{eq}^\mathrm{pl}$ is used to account for isotropic hardening. 
Note that no assumptions on the interactions between $\makro{I}_1$ and $\makro{\varepsilon}_\mathrm{eq}^\mathrm{pl}$ are required. Rather, such interactions, which correspond to distortional hardening effects, can be \enquote{learned} from the training data.
Thirdly, the plastic shear strain $\makro\varepsilon_{12}^\mathrm{pl}$ {(with respect to the coordinate system being aligned with the lattice vectors as shown in \figurename~\ref{fig:Theorie_Homogen_MikroMakro})} is taken as an input to account for an evolving anisotropy of the cubic microstructure under consideration. 
The absolute value $\abs{\makro\varepsilon_{12}^\mathrm{pl}}$ is employed due to the mirror symmetry of the microstructure. {An initial anisotropy of the yield surface could be incorporated by adding $\makro\sigma_{12}$ to the input parameters of $\NNfliess$, but is negligible for the subsequent applications.}
The rate of equivalent plastic strain is identified with the Frobenius norm of the rate of plastic deformation
\begin{equation}
	\Aenderung{\makro{\varepsilon}}_\mathrm{eq}^\mathrm{pl} = \Fnorm{\Aenderung{\makro{\varepsilon}}_{ij}^\mathrm{pl}}\,.
	\label{eq:eqplaststrainrate}
\end{equation}
The flow rule is formulated as
\begin{equation}
	\Aenderung{\makro\varepsilon}_{ij}^\mathrm{pl} 
	=
	\makro\plasMulti \makro{n}_{ij}^\text{pl}
	=
	\Aenderung{\makro{\varepsilon}}_\mathrm{eq}^\mathrm{pl} \underbrace{\frac{\makro{n}_{ij}^\text{pl}}{\Fnorm{\makro{n}_{kl}^\text{pl}}}}_{=:\makro{\breve{n}}_{ij}^\text{pl}}
	\label{equ:flowrule}
\end{equation}
since associated yielding at the micro-scale of elastic-plastic material does not necessarily imply macroscopic associative flow. 
The macroscopic plastic multiplier $\makro\plasMulti$ cannot be extracted from the RVE simulations, but only the plastic strain. This is the reason why the flow rule~\eqref{equ:flowrule} is re-scaled to the rate of the equivalent plastic strain $\Aenderung{\makro{\varepsilon}}_\mathrm{eq}^\mathrm{pl}$.
{It is not reasonable to represent the normalized direction of flow $\makro{\breve{n}}_{ij}^\mathrm{pl}$ directly by a neural network. Firstly, the condition of having unit length $\makro{\breve{n}}_{ij}^\mathrm{pl}\makro{\breve{n}}_{ij}^\mathrm{pl}=1$ is a side condition. Secondly, $\makro{\breve{n}}_{ij}^\mathrm{pl}$ cannot depend only on invariants, but it depends on the complete stress tensor $\makro\sigma_{ij}$ even for isotropic material. Here, the assumption is taken that the principal axes of $\makro{\breve{n}}_{ij}^\mathrm{pl}$ always coincide with those of $\makro\sigma_{ij}$. What remains open is the ratio between the deviatoric and spherical parts of $\makro{\breve{n}}_{ij}^\mathrm{pl}$. This ratio is quantified within the present study by an angle $\dilatflowangle$ within the space $I_1(\makro{\breve{n}}_{ij}^\mathrm{pl})$ -- $J_2(\makro{\breve{n}}_{ij}^\mathrm{pl})$, which results in a flow direction\footnote{Equation~\eqref{equ:MakElasPlas_MatMod_hybrid_Fliessrichtung} is applied only to the in-plane components $i=1..2$, $j=1..2$ as no distinction between plain strain and plane stress is required at the macroscopic scale as explained in Section~\ref{sec:elasticplasticmacro}.}
\begin{equation}
	\makro{\breve{n}}_{ij}^\mathrm{pl}=\frac12\left(\sin\dilatflowangle+\cos\dilatflowangle\right)\delta_{ij}+\left(\cos\dilatflowangle-\sin\dilatflowangle\right)\frac{\mathrm{dev}(\makro\sigma_{ij})}{2\sqrt{J_2(\makro\sigma_{kl})}}\,.
	\label{equ:MakElasPlas_MatMod_hybrid_Fliessrichtung}
\end{equation}
The angle $\dilatflowangle$ of dilatational flow is represented by a second neural network
\begin{equation}
	\dilatflowangle=\NNdirec\left(\makro{I}_1,\makro\varepsilon_\mathrm{eq}^\mathrm{pl},\abs{\makro\varepsilon_{12}^\mathrm{pl}}\right)\,,
	\label{equ:MakElasPlas_MatMod_hybrid_Fliesswinkel}
\end{equation}
}
with the same functional dependencies as the yield function~\eqref{equ:MakElasPlas_MatMod_hybrid_Fliessfunk}. 

Traditionally, macroscopic yielding of porous materials is identified with a limit load, e.~g.\ \citep{Gibson1989}. In this case, it can be shown that macroscopic yielding is associative, cf.~\citep{Leblond2018CRMechanique} and references therein. Usually, limit load analyses are used for analytical estimates with rigid-plastic models.
For the present elastic-plastic model, the macroscopic dissipation amounts to
\begin{equation}
	\Aenderung{\makro\Dissi} = \makro{\sigma}_{ij}\, \Aenderung{\makro{\varepsilon}}_{ij}^\mathrm{pl}\,.
	\label{equ:MakElasPlas_MatMod_AendDissi}
\end{equation}
Since the left-hand side should be equal to the average microscopic dissipation according to Eq.~\eqref{equ:dissipationmicromacro}, macroscopic yielding $\makro{\Fliessfunk}=0$ has to be identified with any yielding at the micro-scale. 
This means that the equivalent yield stress function $\NNfliess$ in Eq.~\eqref{equ:MakElasPlas_MatMod_hybrid_Fliessfunk} covers the whole transition from the first microscopic plastification to complete plastification of the foam cell {at the limit load}.

Regarding the dissipation it should be mentioned that the present formulation, Eqs.~\eqref{equ:MakElasPlas_MatMod_hybrid_Fliessfunk} and \eqref{equ:MakElasPlas_MatMod_hybrid_Fliessrichtung}, does not ensure thermodynamic consistency $\Aenderung{\makro\Dissi}\ge0$ ad hoc. However, the present model is trained only with consistent data $\Aenderung{\makro\Dissi}\ge0$, according to Eq.~\eqref{equ:dissipationmicromacro}. The values of $\Aenderung{\makro\Dissi}$ are monitored during the simulations to ensure that {the} model is operating only within well-trained regions. For all subsequent simulations, $\Aenderung{\makro\Dissi} \ge 0$ holds all the time.

{Note that the yield condition~\eqref{equ:MakElasPlas_MatMod_hybrid_Fliessfunk} can be applied to all plane problems. The information, whether it is a plane stress or a plain strain situation is contained only in the training data for $\NNfliess$. For an extension to 3D problems, the third invariant could be added to the list of arguments of $\NNfliess$ and of $\NNdirec$, if considered relevant, and $\makro\varepsilon_{12}^\mathrm{pl}$ has to be extended towards all respective components of the 3D tensor of plastic strains.}

\subsection{{Elastic-plastic behavior with damage}}
\label{sec:elasticplasticdamagemacro}

Within a second step, the macroscopic model {from Section~\ref{sec:elasticplasticmacro} {is} extended towards damage in order to describe the material behavior with material degradation at the microscale.}
The main step to do so within the present hybrid approach is to identify suitable internal state variables for which values can be extracted from RVE simulations for training. Firstly, it can be stated that damage can evolve anisotropically within the microstructure (although the damage in the bulk material is isotropic, Section~\ref{sec:damagemicro}). 
Secondly, damage $\mikro{\Damage}$ at the microscale degrades the stiffness tensor, which will result in degradation of the macroscopic elastic stiffness $\makro C_{ijkl}$. Current values of $\makro C_{ijkl}$ can be extracted from the RVE simulations by partial unloading. That is why $\makro C_{ijkl}$ is taken as internal state variable, cf.\ e.~g.\ \citep{Wulfinghoff2017IntJSolidsStruct}. Furthermore, it is assumed that $\makro C_{ijkl}$ evolves with the equivalent plastic strain {and that degradation is rate independent} as it is the case at the micro-scale in Eq.~\eqref{equ:Theorie_Schaedigungsentw}:
\begin{equation}
	\Aenderung{\makro{C}}_{ijkl} = \NNfunc^{\Aenderung{\makro{C}}}_{ijkl}\left(\makro{I}_1,\makro\varepsilon_\mathrm{eq}^\mathrm{pl},\abs{\makro\varepsilon_{12}^\mathrm{pl}}\right)\,\Aenderung{\makro\varepsilon}_\mathrm{eq}^\mathrm{pl}\,.
	\label{equ:MakElasPlas_MatMod_Dissi_C}
\end{equation}
The particular coefficients of the evolution equation~\eqref{equ:MakElasPlas_MatMod_Dissi_C} are represented by an additional trained neural network $\NNfunc^{\Aenderung{\makro{C}}}_{ijkl}$.
The macroscopic state law~\eqref{equ:statelawmacro} corresponds to a \textsc{Helmholtz} free energy of
 \begin{equation}
 	\makro\innerEner = \tfrac{1}{2}\,\left(\makro\varepsilon_{ij} - \makro\varepsilon_{ij}^\mathrm{pl}\right)\makro{C}_{ijkl}\left(\makro\varepsilon_{kl} - \makro\varepsilon_{kl}^\mathrm{pl}\right)\,.
 	\label{equ:MakElasPlas_MatMod_hybrid_innerEner}
 \end{equation}
With $\makro C_{ijkl}$ as additional state variable, the dissipation thus amounts to
\begin{equation}
	\Aenderung{\makro\Dissi} 
	= 
	\makro{\sigma}_{ij}\, \Aenderung{\makro{\varepsilon}}_{ij}^\mathrm{pl} -
	\frac{\partial \makro\innerEner}{\partial \makro{C}_{ijkl}} \Aenderung{\makro{C}}_{ijkl} 
	=
	\makro{\sigma}_{ij}\, \Aenderung{\makro{\varepsilon}}_{ij}^\mathrm{pl} -
	\frac{1}{2}\left(\makro\varepsilon_{ij} - \makro\varepsilon_{ij}^\mathrm{pl}\right)\NNfunc^{\Aenderung{\makro{C}}}_{ijkl}\left(\makro\varepsilon_{kl} - \makro\varepsilon_{kl}^\mathrm{pl}\right) \Aenderung{\makro\varepsilon}_\mathrm{eq}^\mathrm{pl}\,.
	\label{equ:MakElasPlas_MatMod_AendDissi_Damage}
\end{equation}
Again, $\Aenderung{\makro\Dissi}\ge0$ is not guaranteed a priori. Anyway, a negative semi-definite coefficient tensor $\NNfunc^{\Aenderung{\makro{C}}}_{ijkl}$ ensures $\Aenderung{\makro\Dissi}\ge0$. This condition will be checked for subsequent applications.

\subsection{Neural network}

\begin{figure}
	\centering
	\includegraphics[width=0.7\textwidth]{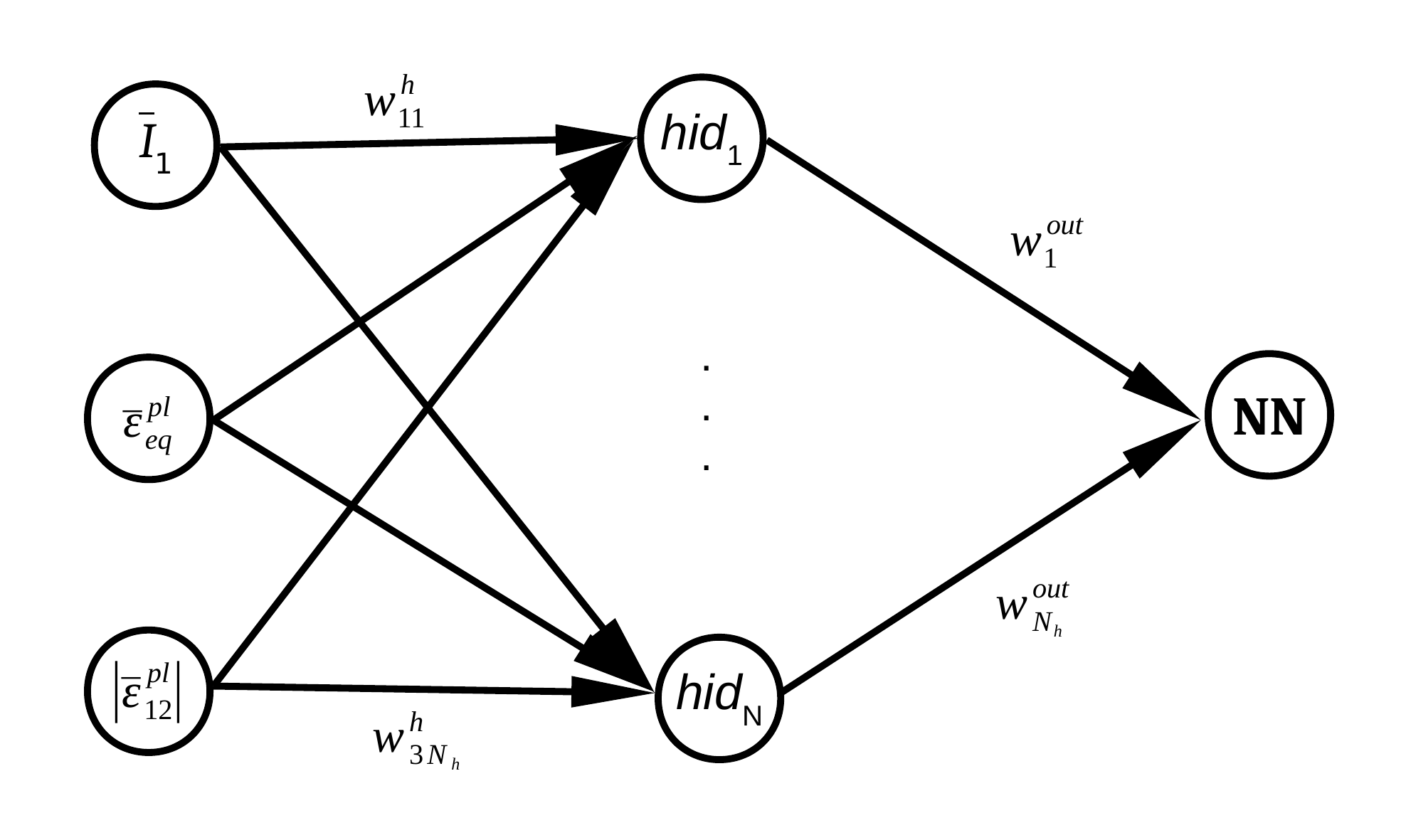}
	\caption{Structure of the employed neural network: input layer with $\inpNeuron$ variables, a single hidden layer with $\hidNeuron$ neurons and output layer with $\outNeuron$ neurons}
	\label{fig:TheorieNN_strukt}
\end{figure}

{The two constitutive functions $\NNfliess$ and $\NNdirec$ for the elastic-plastic model, Section~\ref{sec:elasticplasticmacro}, and the three functions $\NNfliess$, $\NNdirec$ and $\NNfunc^{\Aenderung{\makro{C}}}_{ijkl}$ }
for yield condition, flow rule and evolution of damage, respectively, are represented each in Voigt notation by a neural network with single hidden layer 
\begin{equation}
	\NNfunc_k(\mathrm{inp}_i)=\sum_{j=1}^{\hidNeuron} \Weight^{\mathrm{out}}_{kj}g\left(\sum_i\Weight^{\mathrm{h}}_{ji}\mathrm{inp}_i+\Bias^{\mathrm{h}}_j\right)+\Bias^{\mathrm{out}}_k
\end{equation}
having $\hidNeuron=100$ neurons as shown schematically in \figurename~\ref{fig:TheorieNN_strukt}. 
The sigmoid function $g(x)=\frac{1}{1+\exp(-x)}$ is employed as activation function.
The mean squared error
\begin{equation}
	err = \tfrac{1}{2}\sum_{m\in \trainData}\sum_k\left[val^{(m)}_k-\NNfunc_k(\mathrm{inp}_i^{(m)})\right]^2
	\label{equ:TheorieNN_ErrFuncDefault}
\end{equation}
between training data $\left\lbrace val^{(m)}_k, \mathrm{inp}^{(m)}_i=\left(\makro{I}_1,\makro\varepsilon_\mathrm{eq}^\mathrm{pl},\abs{\makro\varepsilon_{12}^\mathrm{pl}}\right)^{(m)}\right\rbrace$ and respective output of the neural network is minimized \emph{offline} by a truncated Newton method (with normalized data) to determine the weights $\Weight^{\mathrm{out}}_{kj}$, $\Weight^{\mathrm{h}}_{ji}$ and biases $\Bias^{\mathrm{out}}_k$, $\Bias^{\mathrm{h}}_j$. 
The generated data $\Datenmenge$ are split into subsets for training $\trainData$ and for validation $\valiData$, whereby the training set encompasses about 90\% of the data.

\subsection{Generation of training data}
\label{sec:training}

The sets of training data $\Datenmenge$ for the elastic-plastic model {and for the elastic-plastic model with damage} are generated by RVE simulations. {Proportional loading paths in the macroscopic strain space}
\begin{equation}
		\makro{\varepsilon}_{ij} = \proFak \makro{\varepsilon}_{ij}\init
		\label{equ:MakElasPlas_MatMod_propor}
\end{equation}
{are employed for simplicity}
with a given direction $\makro{\varepsilon}_{ij}\init$ and a time-like factor of proportionality $\proFak$.
The plane problem under consideration has three independent strain components $\makro{\varepsilon}_{11}$, $\makro{\varepsilon}_{22}$, $\makro{\varepsilon}_{12}$, respectively. 
Equivalently, a strain state can be characterized by principal strains $\makro{\varepsilon}_{I}$ and $\makro{\varepsilon}_{II}$, together with the angle $\Psi=\frac{1}{2} \arctan\left({2\makro{\varepsilon}_{12}}/({\makro{\varepsilon}_{11}-\makro{\varepsilon}_{22}})\right)$ between principal directions and the coordinate axes.  
Consequently, the direction $\makro{\varepsilon}_{ij}\init$ can be characterized by two parameters $\phi=\arccos\left(\makro{\varepsilon}_{I}/{\sqrt{\makro{\varepsilon}_{I}^2+\makro{\varepsilon}_{II}^2}}\right)$ and $\Psi$, respectively. Both \enquote{angles} are rasterized with increments $\Delta\phi=\ang{10}$, $\Delta\Psi=\ang{10}$ over the $\phi$--$\Psi$ space for generation of $\Datenmenge$.  Subsequently, the direction tensor $\makro{\varepsilon}_{ij}\init$ is re-scaled such that $\proFak=1$ corresponds to initial yielding. {In total, 324 RVE simulations are performed, and about 150 time increments from each simulation are used for the training.}

Macroscopic yielding $\makro{\Fliessfunk}=0$ is defined as active plastic deformations at the micro-scale as discussed in Section~\ref{sec:elasticplasticmacro}. In view of the structure of the yield function in Eq.~\eqref{equ:MakElasPlas_MatMod_hybrid_Fliessfunk}, this means that
	\begin{equation}
    \NNfliess\left(\makro{I}_1,\makro\varepsilon_\mathrm{eq}^\mathrm{pl},\abs{\makro\varepsilon_{12}^\mathrm{pl}}\right)=\Fnorm{\makro\sigma_{ij}}
	\end{equation}
holds. The right-hand side of this equation is known from the RVE simulation and can thus be used as training data for $\NNfliess$. The Frobenius norm  $\Fnorm{\makro\sigma_{ij}}$ of the macroscopic stress tensor is used instead of von Mises equivalent stress to obtain unique training data even for purely hydrostatic states of loading.
The current values of the macroscopic plastic strain $\makro\varepsilon_{kl}^\mathrm{pl}$ can be obtained by inversion of the state law~\eqref{equ:statelawmacro}:
\begin{equation}
    	\makro\varepsilon_{kl}^\mathrm{pl}  = \makro\varepsilon_{kl}-\makro C^{-1}_{ijkl}\makro\sigma_{ij}\,.
		\label{equ:statelawmacroinvertedplastic}
\end{equation}
The rate of plastic deformation, required as training data for the flow rule~\eqref{equ:flowrule}, can be obtained subsequently by a backward difference quotient
\begin{equation}
	 \Aenderung{\makro{\varepsilon}}_{ij}^\mathrm{pl} \approx
	    (\left.\makro\varepsilon_{kl}^\mathrm{pl}\right|_{t}-\left.\makro\varepsilon_{kl}^\mathrm{pl}\right|_{t-\Delta t})/\Delta t \,.
\end{equation}
This value is used to extract the rate of equivalent plastic strain $\Aenderung{\makro{\varepsilon}}_\mathrm{eq}^\mathrm{pl}$ by Eq.~\eqref{eq:eqplaststrainrate} and subsequently the normalized flow direction $\makro{\breve{n}}_{ij}^\mathrm{pl}=\Aenderung{\makro{\varepsilon}}_{ij}^\mathrm{pl}/\Aenderung{\makro{\varepsilon}}_\mathrm{eq}^\mathrm{pl}$ as training data for $\NNdirec$.

For the elastic-plastic model from Section~\ref{sec:elasticplasticmacro}, the constant macroscopic stiffness $\makro C_{ijkl}$ in state law~\eqref{equ:statelawmacro} is computed \emph{offline} by a conventional linear perturbation analysis, 
{ and $\NNfliess$ and $\NNdirec$ are trained from data which are extracted from RVE simulations without microscopic damage.}
For the {elastic-plastic model with damage} from Section~\ref{sec:elasticplasticdamagemacro}, current values of $\makro C_{ijkl}$ are obtained by partial unloading in different strain directions. In particular, a first proportional unloading step of $\Delta\makro{\varepsilon}_{ij}=-0.025\makro{\varepsilon}_{ij}$ in direction of the current loading is performed to ensure that the material is completely within the elastic domain. Subsequently, unloading steps of additional 2.5\% are applied in all (independent) directions of $\makro{\varepsilon}_{ij}$ to extract the current stiffness $\makro{C}_{ijkl}$.
Again, a difference quotient is applied to compute the rate $\Aenderung{\makro{C}}_{ijkl}$ as training data for $\NNfunc^{\Aenderung{\makro{C}}}_{ijkl}$ in Eq.~\eqref{equ:MakElasPlas_MatMod_Dissi_C},
{together with $\NNfliess$ and $\NNdirec$ from RVE simulations with microscopic damage.}

\subsection{Implementation}

The macroscopic material model was implemented as a user-defined material routine into the commercial FEM code Abaqus/Standard via the UMAT interface. The update of the internal state variables is done by means of the regula falsi method as outlined in appendix~\ref{app:implementation}. 
Although all required derivatives for the consistent tangent could be computed in principle from the neural network representations, as done e.~g.\ in \citep{Jung2006ComputStruct,Le2015,Liang2008EngStruct}, a forward finite difference scheme is adopted in the present implementation. The computational overhead of about 60-80\% \citep{PerezFoguet2000ComputMethodsApplMechEngrg} is acceptable for the present proof-of-concept study.
The neural networks are implemented using the library FFNET \citep{FFNET,Wojciechowski2011FFNET}. Training is performed \emph{offline} in Python. The response functions can be evaluated efficiently \emph{online} via the Fortran interface of FFNET within the UMAT routine.

\section{Applications}
\label{sec:application}

\subsection{Material point}

\begin{figure}[hbt]
	\centering
	\subfigure[$\makro\varepsilon_{22}=0$]{
		\label{fig:MatMod_1RVE_hybrid_Teilentlast_hydro}
		\includegraphics[width=0.28\columnwidth]{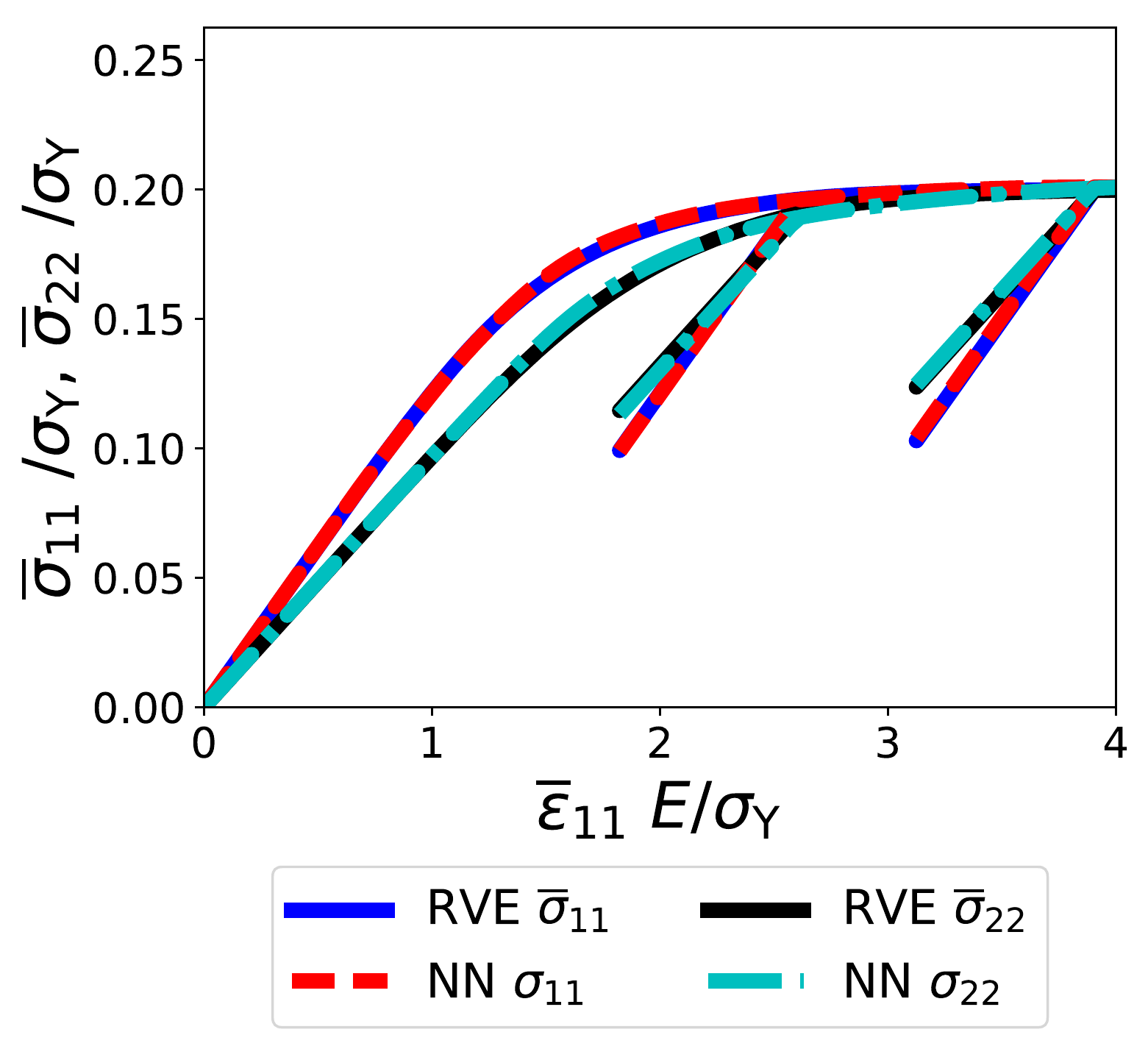} 
	}
	\hfill
	\subfigure[$\makro\varepsilon_{22}=-\makro\varepsilon_{11}$]{
		\label{fig:MatMod_1RVE_hybrid_Teilentlast_shear}
		\includegraphics[width=0.29\columnwidth]{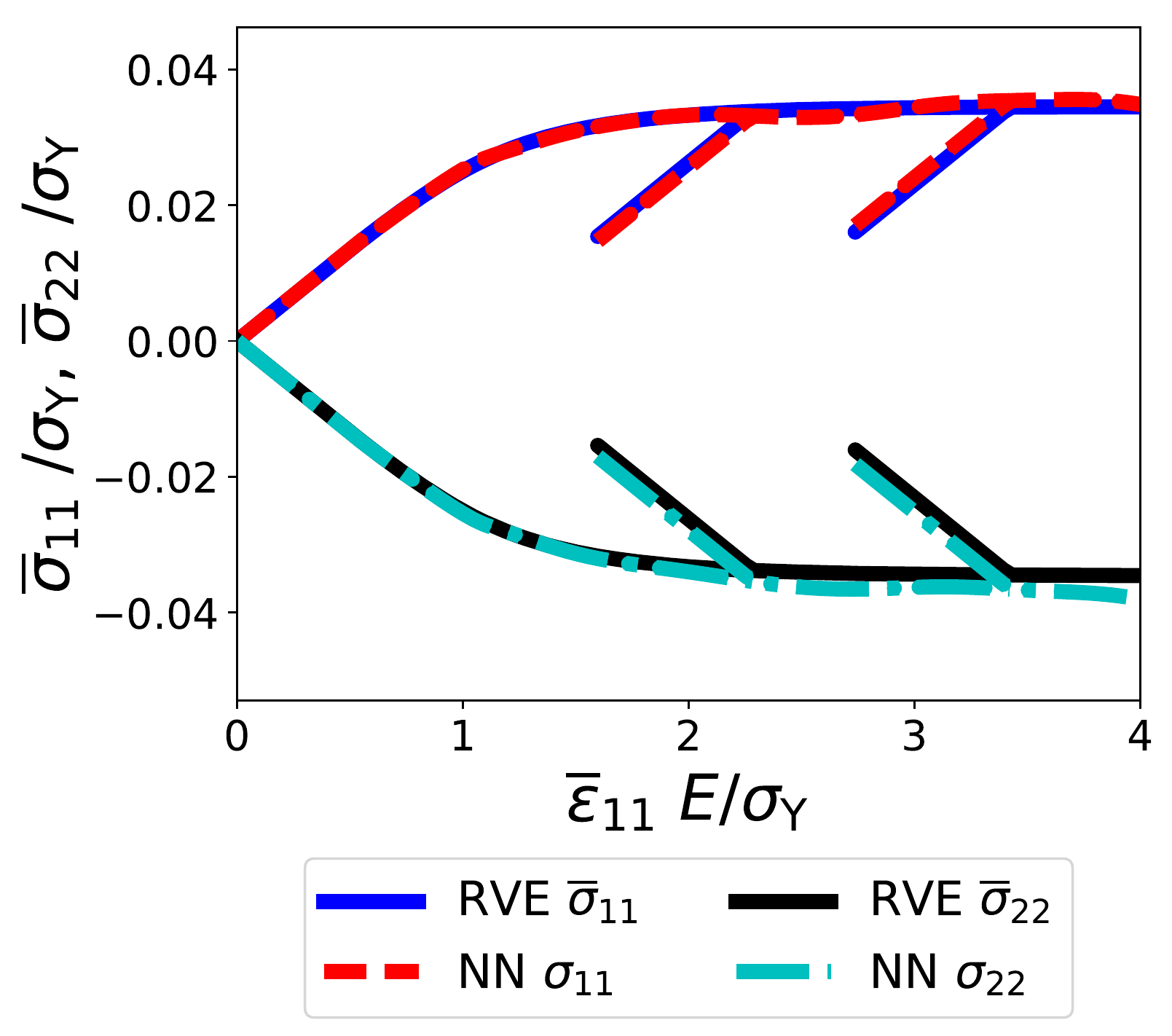}
	}
	\hfill
	\subfigure[$\makro\varepsilon_{22}=-\frac{1}{3}\makro\varepsilon_{11}$]{
		\label{fig:MatMod_1RVE_hybrid_Teilentlast_1over3}
		\includegraphics[width=0.295\columnwidth]{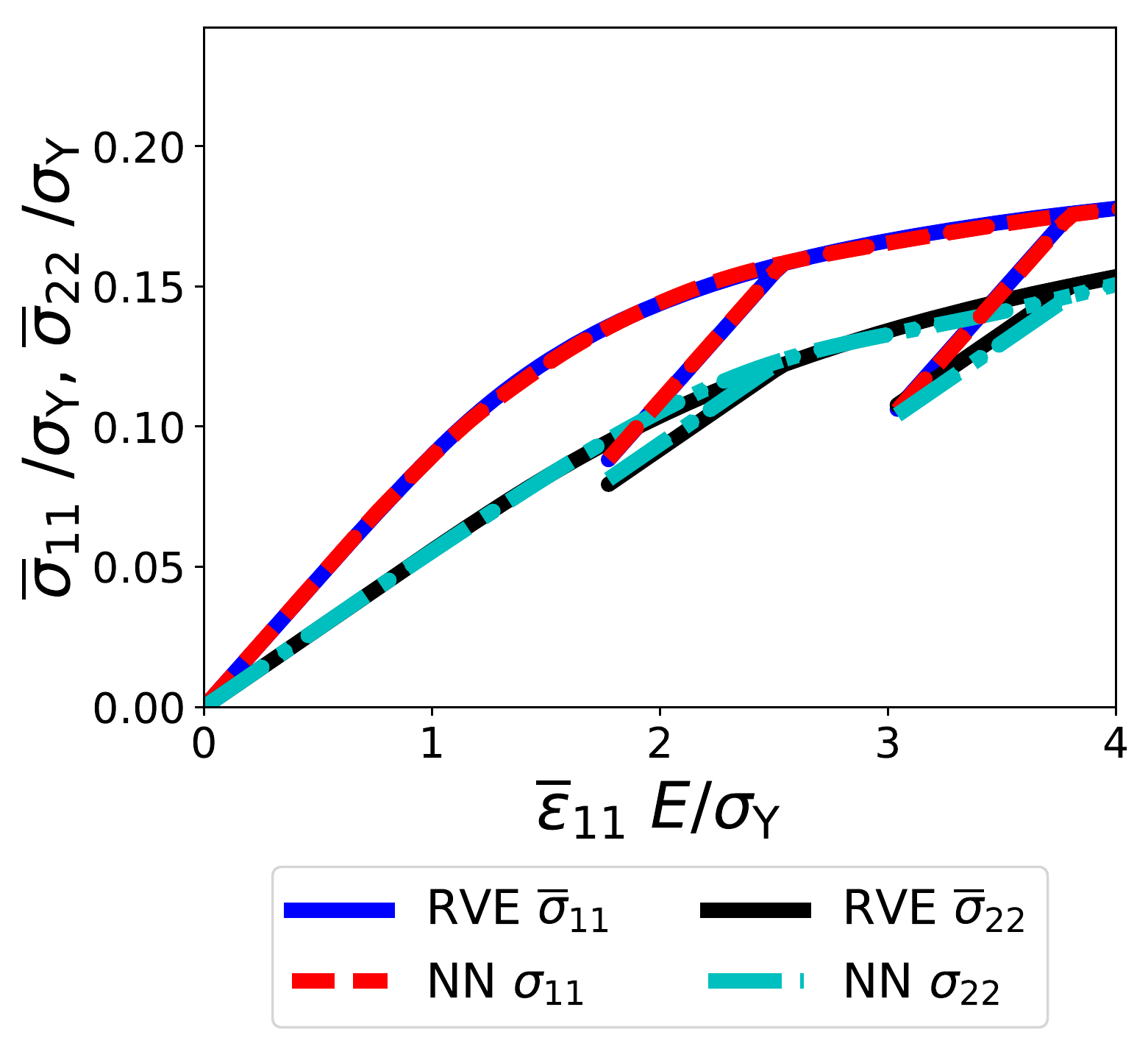}
	}
	\caption{Stress-strain curves of proportional load paths with partial unloading of neural network macroscopic model (\enquote{NN}) in comparison to RVE simulations}
	\label{fig:MatMod_1RVE_hybrid_TeilEntlast}
\end{figure}

Firstly, the performance of the {elastic-plastic} model shall be evaluated for a single material point. Figure~\ref{fig:MatMod_1RVE_hybrid_TeilEntlast} shows the stress-strain responses of the hybrid multi-scale neural network approach (HyMNNA) in comparison to corresponding RVE simulations {without damage (\figurename~\ref{fig:Theorie_Homogen_MikroMakro})} for three proportional loading paths.
For all three loading paths, the envelops of the neural network model and of the RVE simulations agree quite well. This behavior is not surprising, since these envelops have been used for training.
Additionally, \figurename~\ref{fig:MatMod_1RVE_hybrid_TeilEntlast} contains results for partial unloading. Obviously, the present HyMNN approach is able to capture the irreversible behavior in the loading-unloading behavior, in contrast to pseudo-plastic approaches \citep{Settgast2019,Fritzen2019}.

\begin{figure}
	\centering
	\hfill
	\subfigure[]{
	\includegraphics[width=0.47\columnwidth]{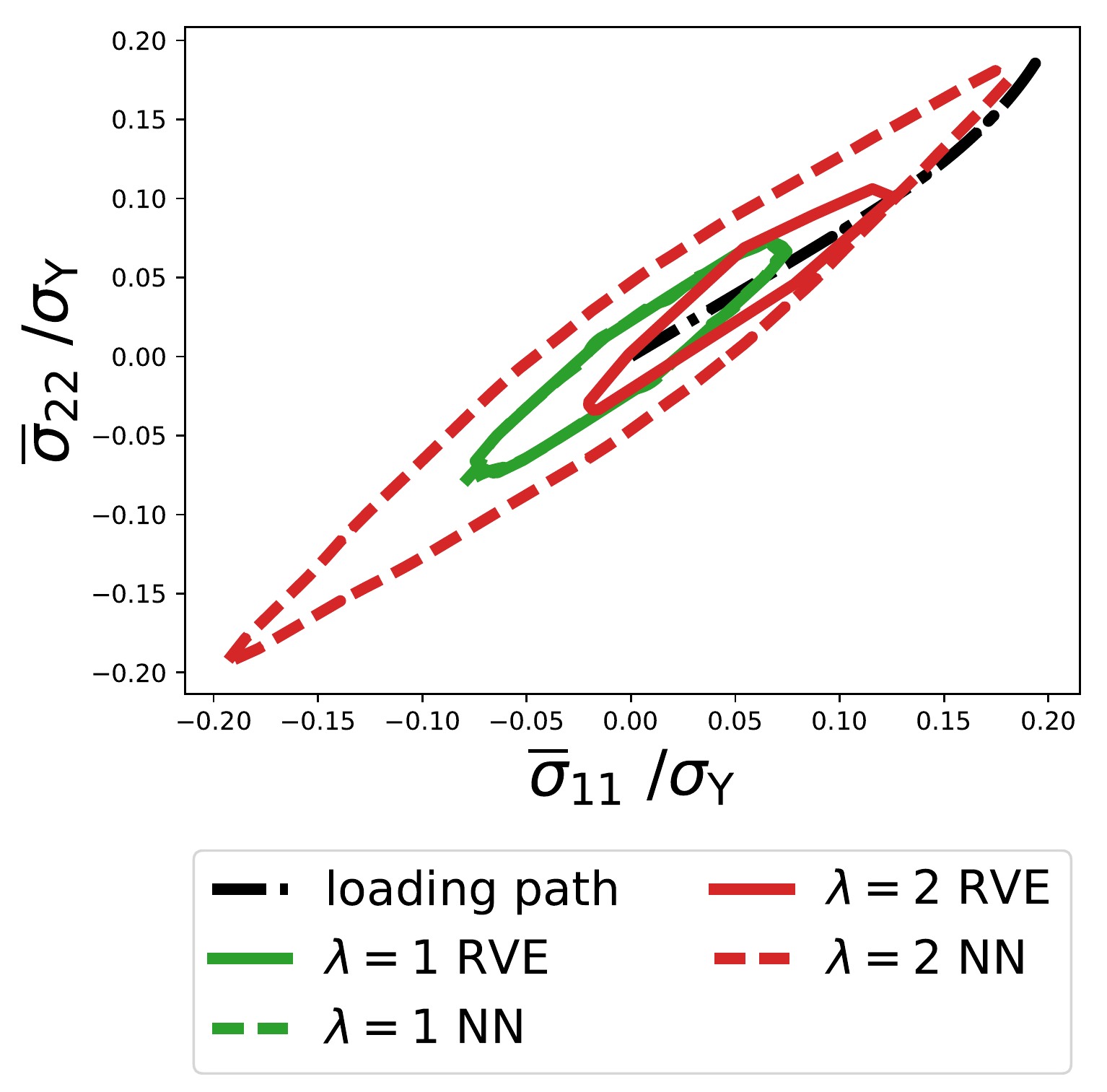}
	\label{fig:MatMod_1RVE_hybrid_TeilEntlast_YieldSurfaces_Evolution}
	}
	\hfill
	\subfigure[]{
	\includegraphics[width=0.47\columnwidth]{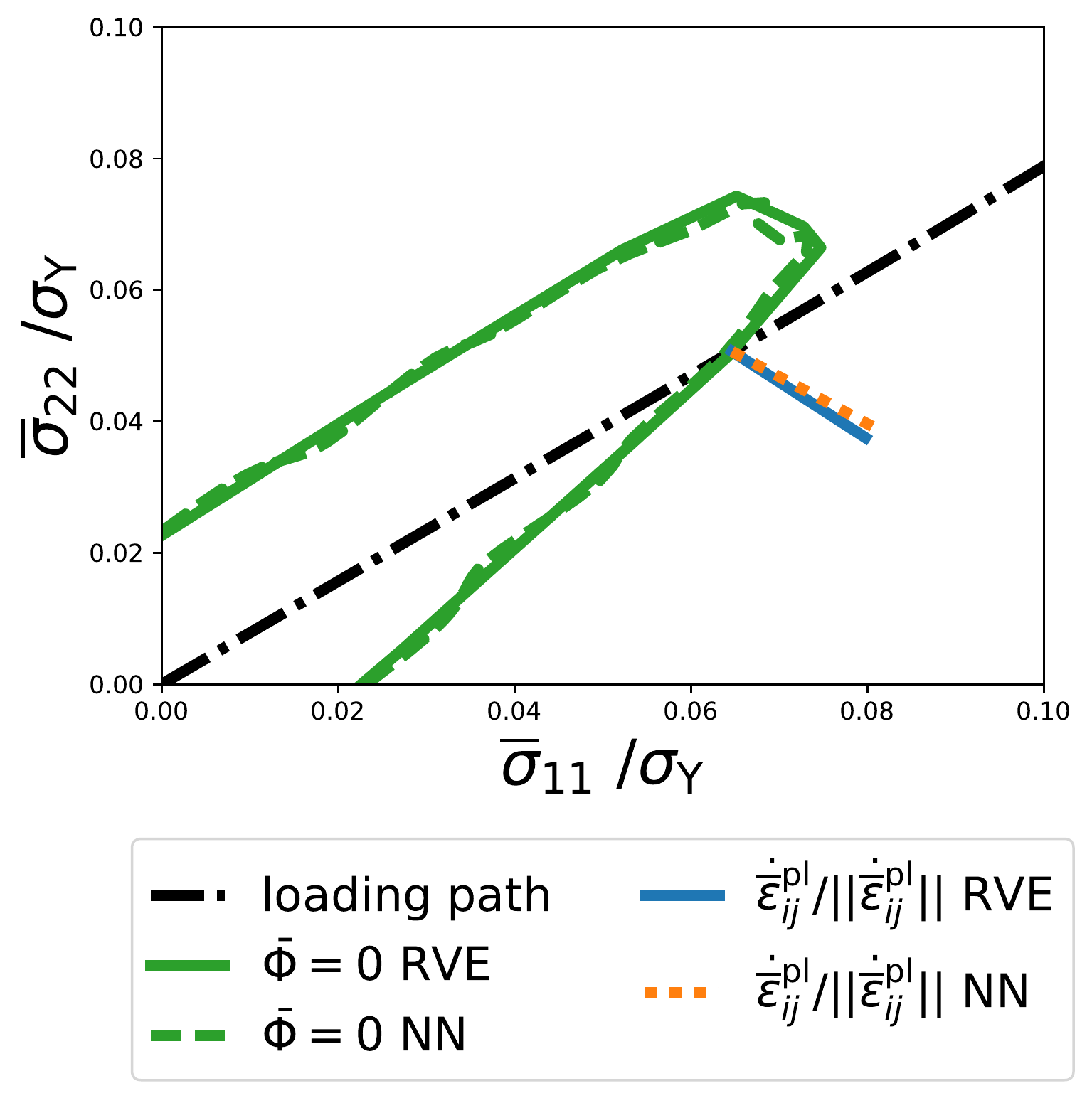}
	\label{fig:MatMod_1RVE_hybrid_TeilEntlast_YieldSurfaces_FlowDir}
	}
	\caption{Yield surface $\makro\Fliessfunk=0$: \subref{fig:MatMod_1RVE_hybrid_TeilEntlast_YieldSurfaces_Evolution} evolution with increasing load $\proFak$, and \subref{fig:MatMod_1RVE_hybrid_TeilEntlast_YieldSurfaces_FlowDir} flow directions for initial yielding $\proFak=1$}
	\label{fig:MatMod_1RVE_hybrid_TeilEntlast_YieldSurfaces}
\end{figure}

\figurename~\ref{fig:MatMod_1RVE_hybrid_TeilEntlast_YieldSurfaces} shows the evolution of the yield surface $\makro\Fliessfunk=0$ for the uni-axial load case from Fig.~\ref{fig:MatMod_1RVE_hybrid_Teilentlast_hydro}, again both for the RVE simulations and for HyMNNA. 
The yield surface was determined numerically by reverse loading in different directions as proposed in \citep{Storm2016IntJMechSci}.
Firstly, it can be seen in \figurename~\ref{fig:MatMod_1RVE_hybrid_TeilEntlast_YieldSurfaces_Evolution} that the yield surface extends mainly in the direction of equi-axial loading $\makro\sigma_{11}=\makro\sigma_{22}$ as it is well-known for foam materials \citep{Gibson1989,Storm2016IntJMechSci}.
The initial yield surface $\proFak=1$ is represented in total quite well by the HyMNN approach. A very
close look at \figurename~\ref{fig:MatMod_1RVE_hybrid_TeilEntlast_YieldSurfaces_FlowDir} shows that the neural network representation of the yield surface exhibits slight oscillations. Although these oscillations are negligible to detect yielding, they would be very relevant if an associated flow rule were used. That is why it is advantageous to represent the flow direction by an independent neural network, Eq.~\eqref{equ:MakElasPlas_MatMod_hybrid_Fliessrichtung}. In this way, the flow direction can be described adequately as can be seen in \figurename~\ref{fig:MatMod_1RVE_hybrid_TeilEntlast_YieldSurfaces_FlowDir}.
\figurename~\ref{fig:MatMod_1RVE_hybrid_TeilEntlast_YieldSurfaces_Evolution} contains additionally the yield surfaces for twice the loading $\proFak=2$. The comparison between both respective yield surfaces shows that the evolution of the yield surface close to the loading path (black dashed line in \figurename~\ref{fig:MatMod_1RVE_hybrid_TeilEntlast_YieldSurfaces}) is captured well.
However, in a certain distance to the current loading path, deviations between the neural network representation and the RVE yield surface are obvious. The RVE yield surface is shifted mainly (kinematic hardening), whereas the neural network yield surface is extended (isotropic hardening).
These deviations are attributed to two reasons. Firstly, only proportional loading paths were used for training, but no loadings in opposite directions. Secondly, {hardening is represented at the macroscopic scale solely by the equivalent plastic strain $\makro\varepsilon_\mathrm{eq}^\mathrm{pl}$, see Section~\ref{sec:elasticplasticmacro}}. 
{Kinematic hardening could be included by incorporating a back stress to the yield condition~\eqref{equ:MakElasPlas_MatMod_hybrid_Fliessfunk}, whose evolution equation can be represented by a neural network as proposed by \citet{Furukawa2004EngAnalBoundElem}. Though, such an approach requires respective training data for the evolution of the back stresses.}
Anyway, the loading paths in the following application are solely monotonic so that the representation by the model from Section~\ref{sec:elasticplasticmacro} is adequate.

\subsection{Damage}

\begin{figure}
	\centering
		\includegraphics{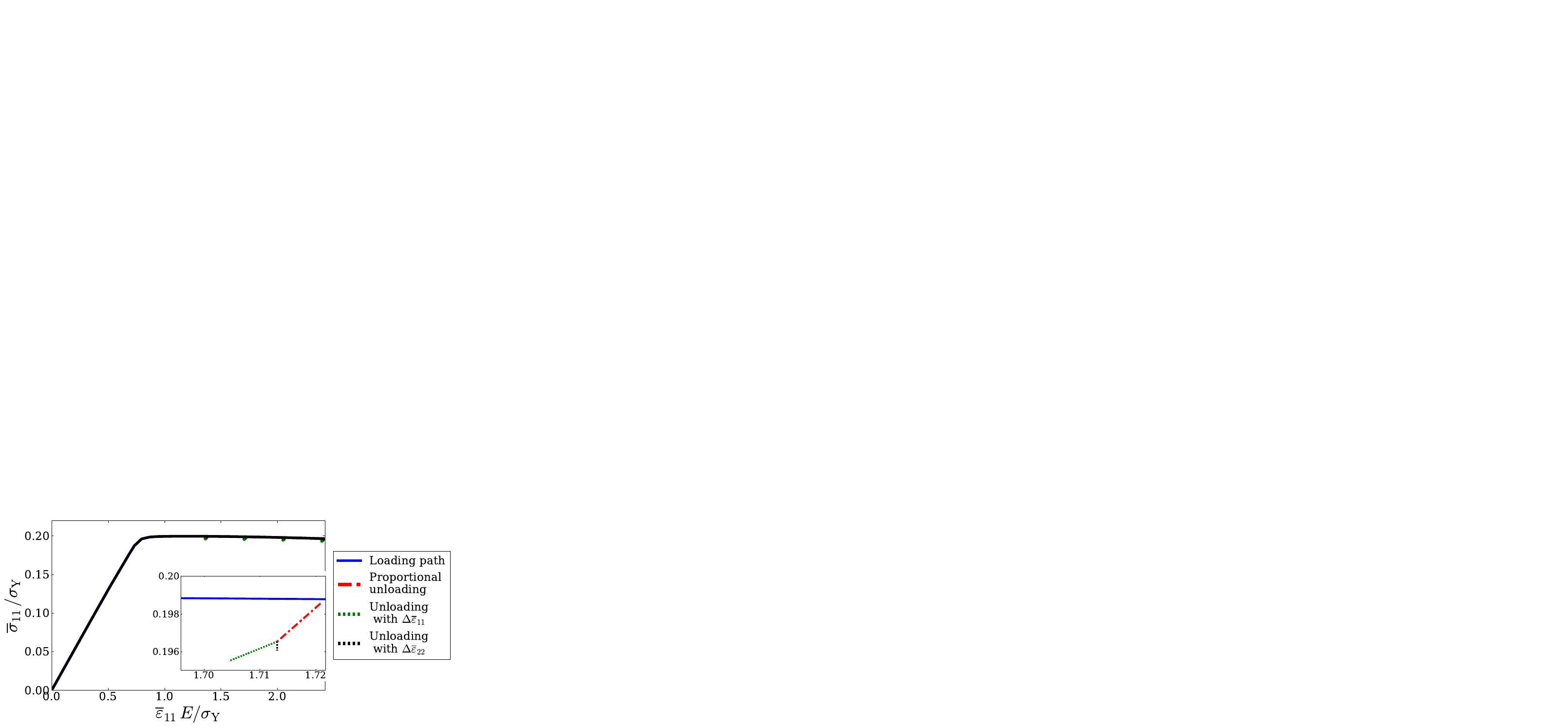}
		\caption{Partial unloading to extract current stiffness}
	\label{fig:MakroElasPlas_Schädigung_ABA_Teilentlast}
\end{figure}

\begin{figure}
	\centering
	\subfigure[uni-axial loading $\makro{\varepsilon}_{22}$, $\makro{\varepsilon}_{11}=\makro{\varepsilon}_{12}=0$ ]{
		\label{fig:MakroElasPlas_Schädigung_ABA_CEinachs}
		\includegraphics[width=0.45\columnwidth]{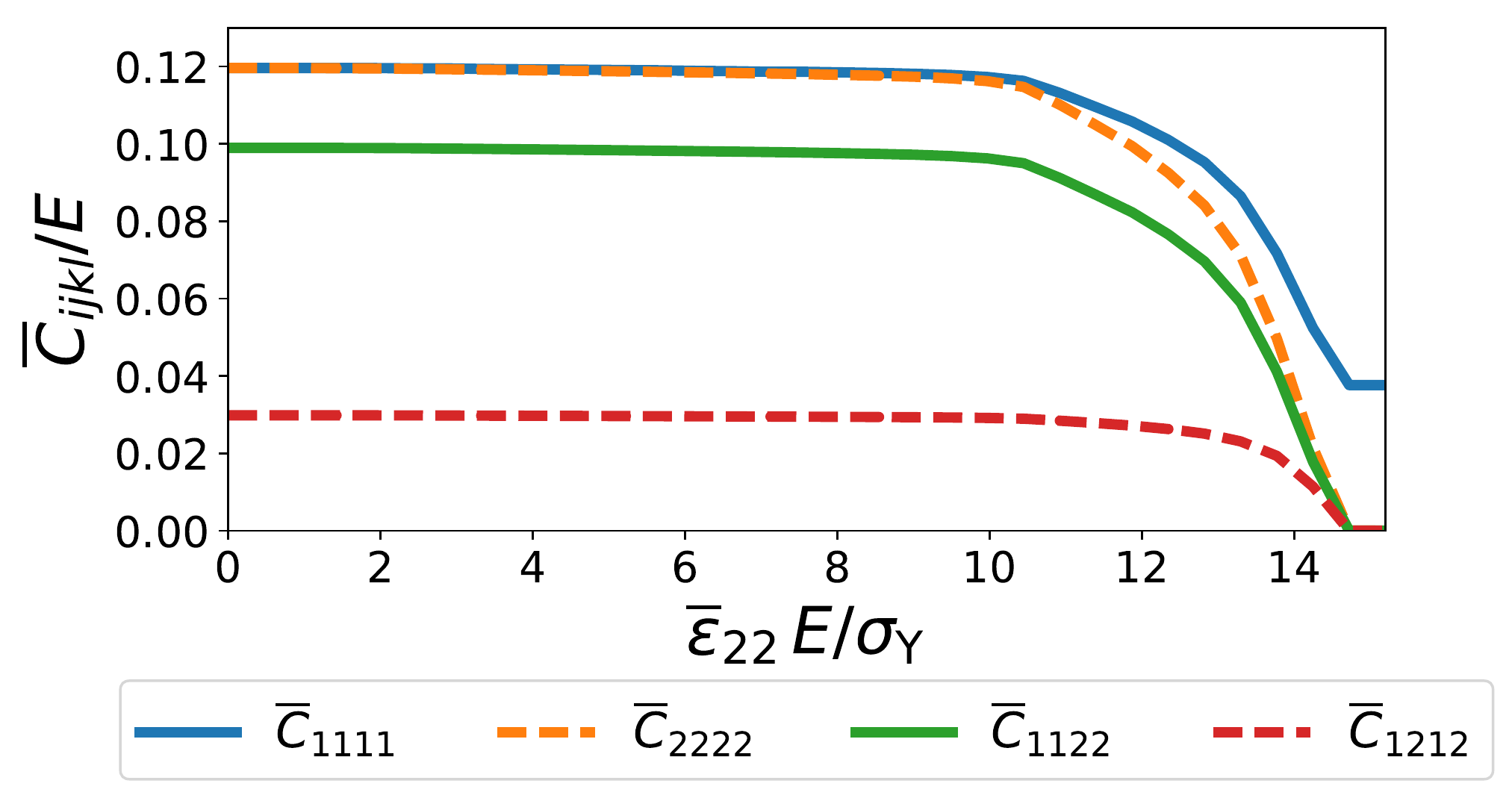}
	}
	\hfill
	\subfigure[bi-axial loading $\makro{\varepsilon}_{11}=\makro{\varepsilon}_{22}$, $\makro{\varepsilon}_{12}=0$]{
		\label{fig:MakroElasPlas_Schädigung_ABA_Chydro}
		\includegraphics[width=0.45\columnwidth]{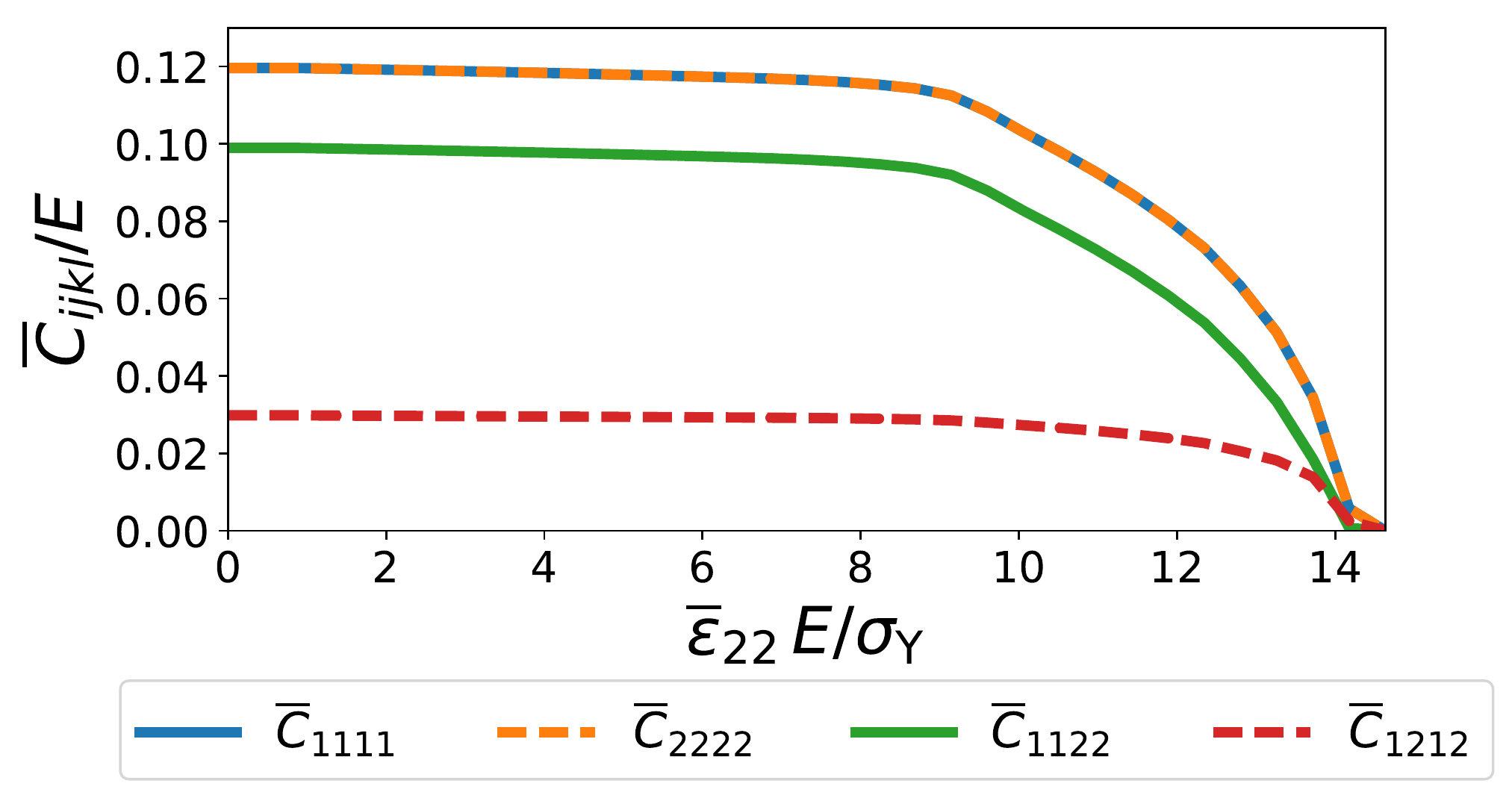}
	}
	\caption{Evolution of macroscopic stiffness for damage evolving at the micro-scale }
	\label{fig:MakroElasPlas_Schädigung_ABA}
\end{figure}

Damage is characterized at the macroscale by degradation of the elastic stiffness tensor~$\makro{C}_{ijkl}$.
\figurename~\ref{fig:MakroElasPlas_Schädigung_ABA_Teilentlast} shows how the current components of $\makro{C}_{ijkl}$ are extracted from the RVE simulations {with microscopic damage} by partial unloading.  
The evolution of these components with increasing loading is plotted in \figurename~\ref{fig:MakroElasPlas_Schädigung_ABA} for two load cases.
Note that components $\makro{C}_{1111}$ and $\makro{C}_{2222}$ evolve differently under uni-axial tension in \figurename~\ref{fig:MakroElasPlas_Schädigung_ABA_CEinachs}, indicating the evolving anisotropy. Even upon complete rupture $\makro{C}_{2222}=0$ at $\makro{\varepsilon}_{11}\approx15\sigma_{\FliessSpa}/E$, the material can still carry load in transversal direction $\makro{C}_{1111}>0$. 
For bi-axial loading in \figurename~\ref{fig:MakroElasPlas_Schädigung_ABA_Chydro}, $\makro{C}_{1111}$ and $\makro{C}_{2222}$ decay simultaneously as expected.

\begin{figure}
	\centering
	\subfigure[uni-axial loading $\makro{\varepsilon}_{22}$, $\makro{\varepsilon}_{11}=\makro{\varepsilon}_{12}=0$, ]{
		\label{fig:MakroElasPlas_Schädigung_NN_Einachs}
		\includegraphics[width=0.295\columnwidth]{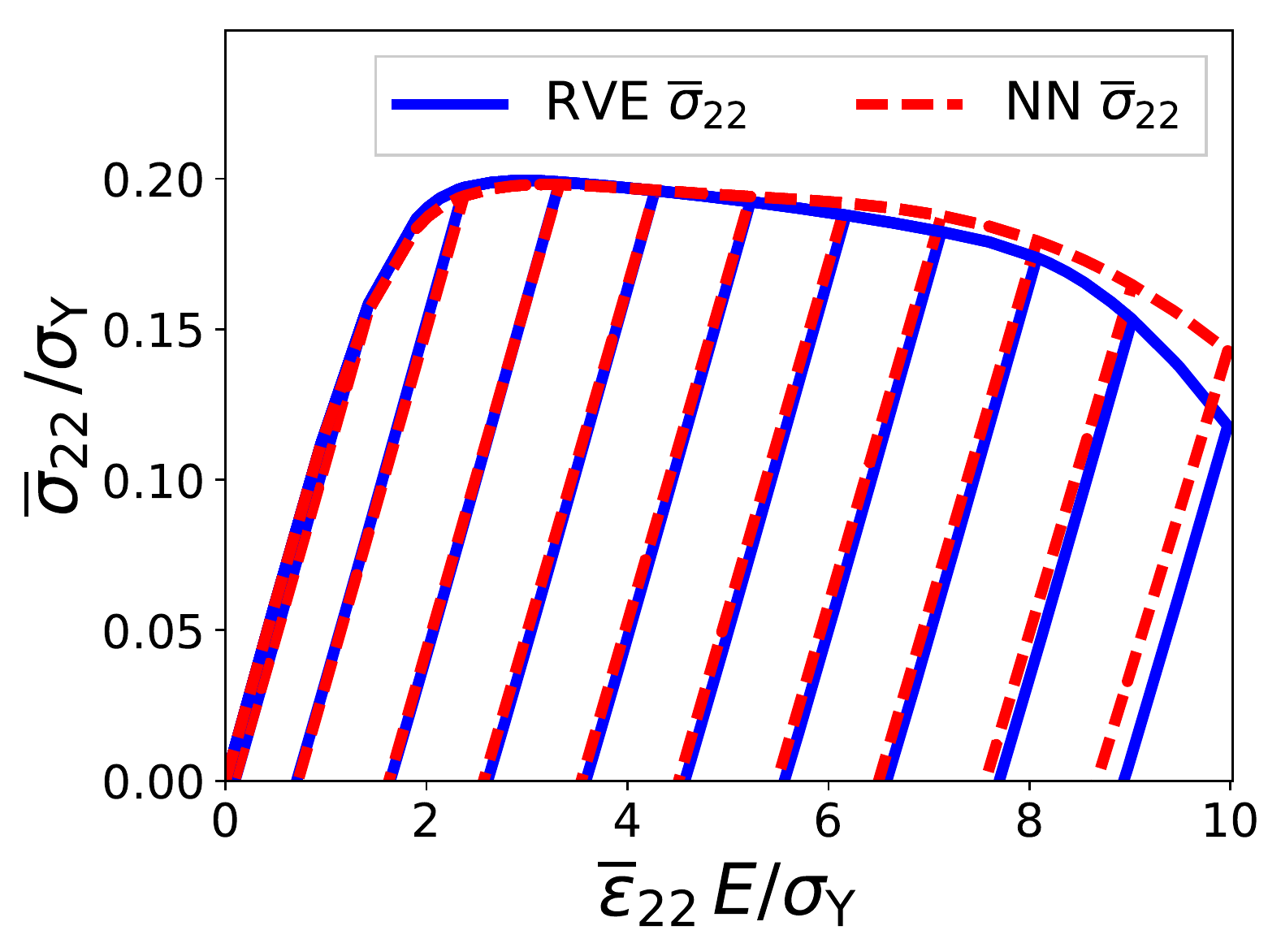}
	}
	\hfill
	\subfigure[bi-axial loading $\makro{\varepsilon}_{11}=\makro{\varepsilon}_{22}$, $\makro{\varepsilon}_{12}=0$]{
		\label{fig:MakroElasPlas_Schädigung_NN_Hydro}
		\includegraphics[width=0.295\columnwidth]{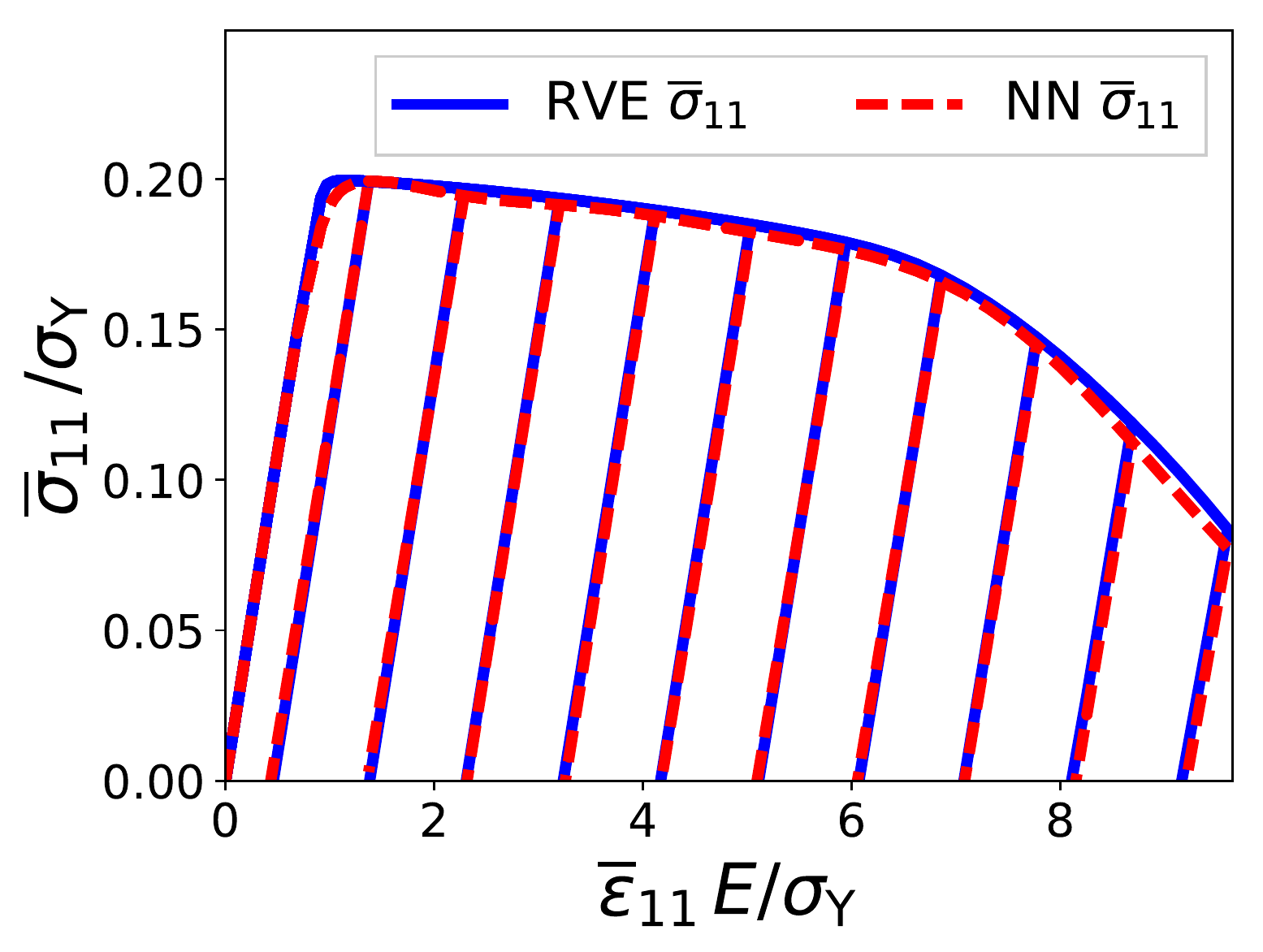}
		\includegraphics[width=0.295\columnwidth]{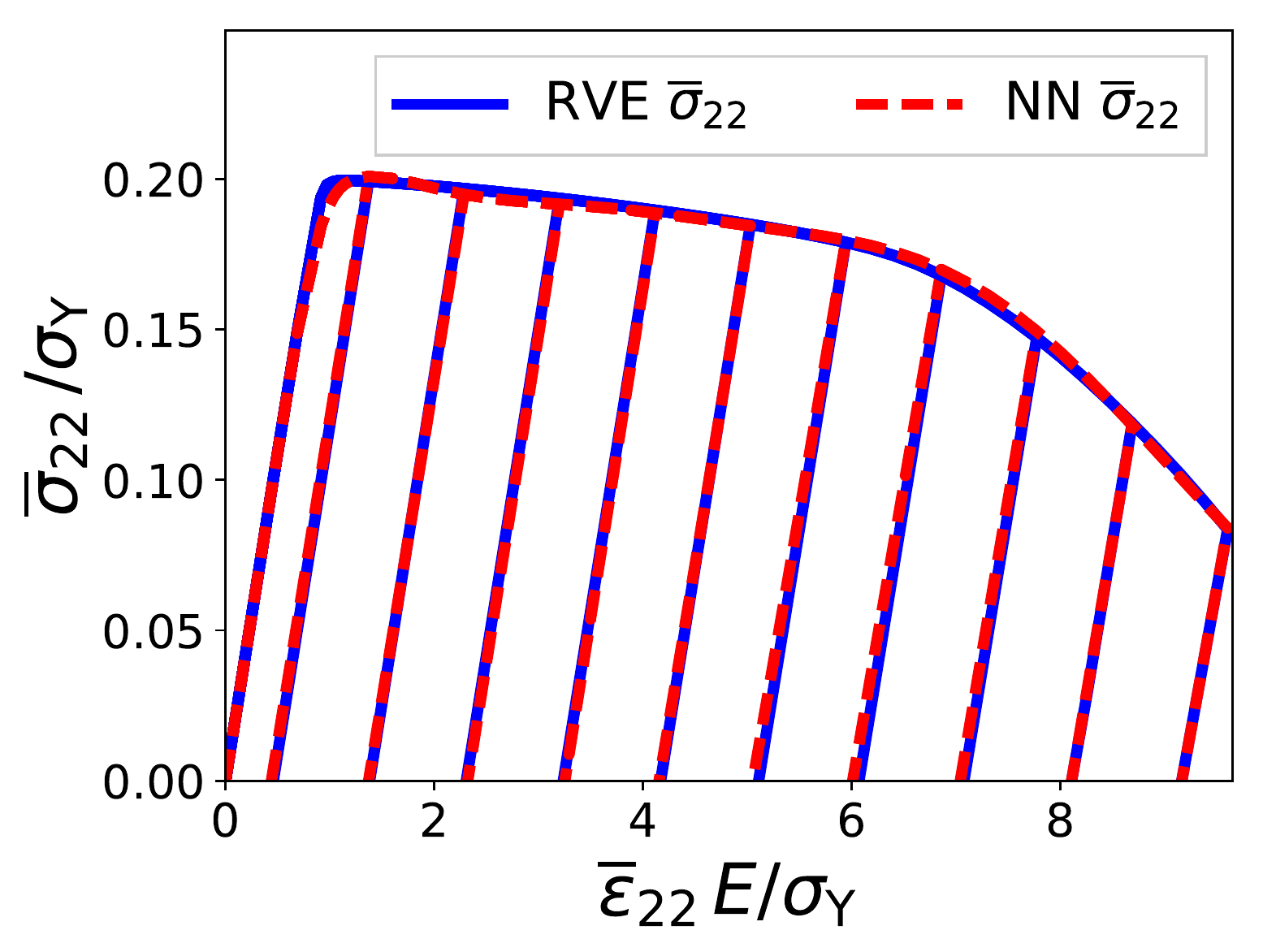}
	}
	\hfill
	\subfigure[deviatoric loading $\makro{\varepsilon}_{11}=-\makro{\varepsilon}_{22}$, $\makro{\varepsilon}_{12}=0$]{
		\label{fig:MakroElasPlas_Schädigung_NN_Devia}
		\includegraphics[width=0.295\columnwidth]{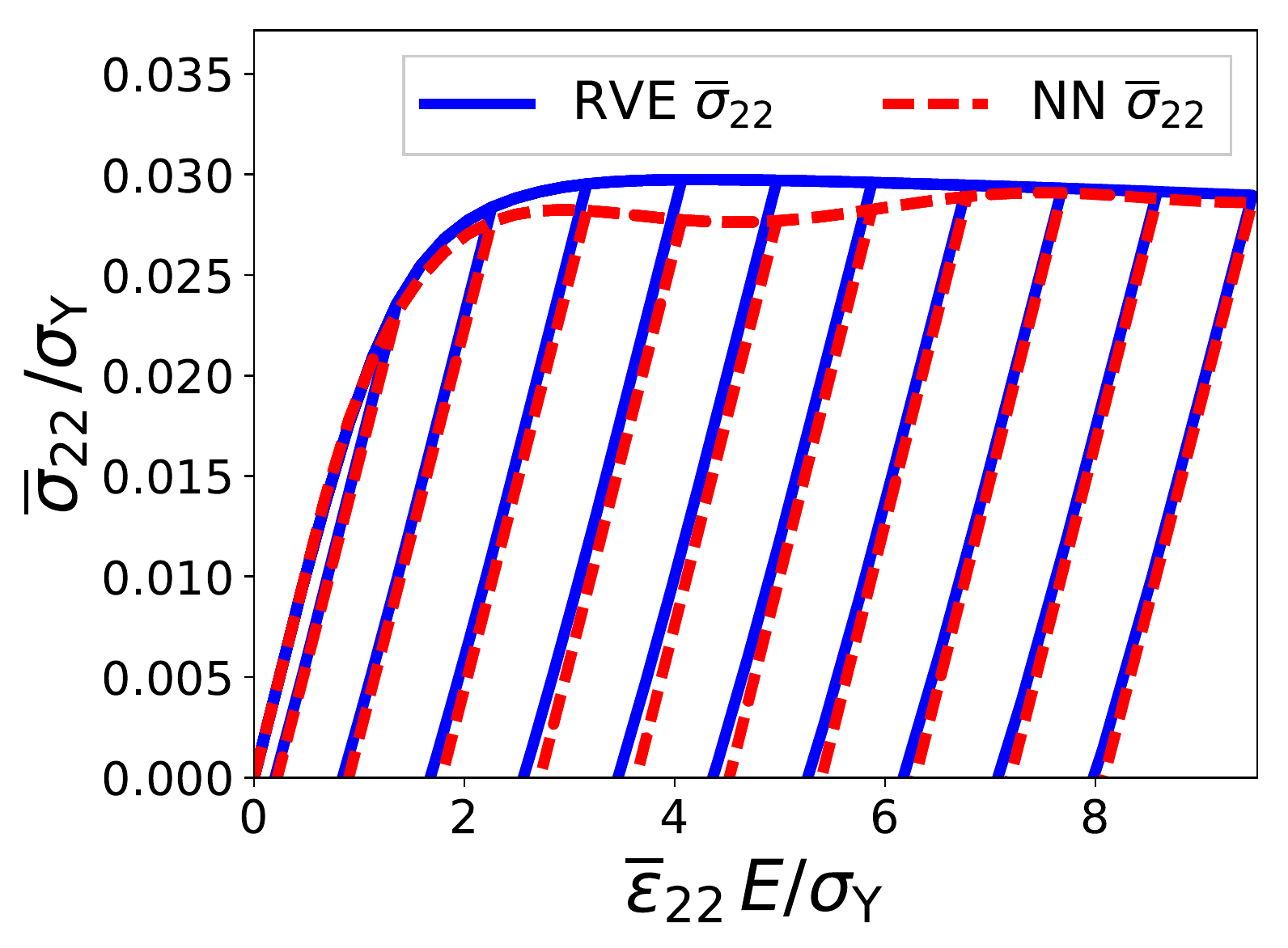}
		\includegraphics[width=0.295\columnwidth]{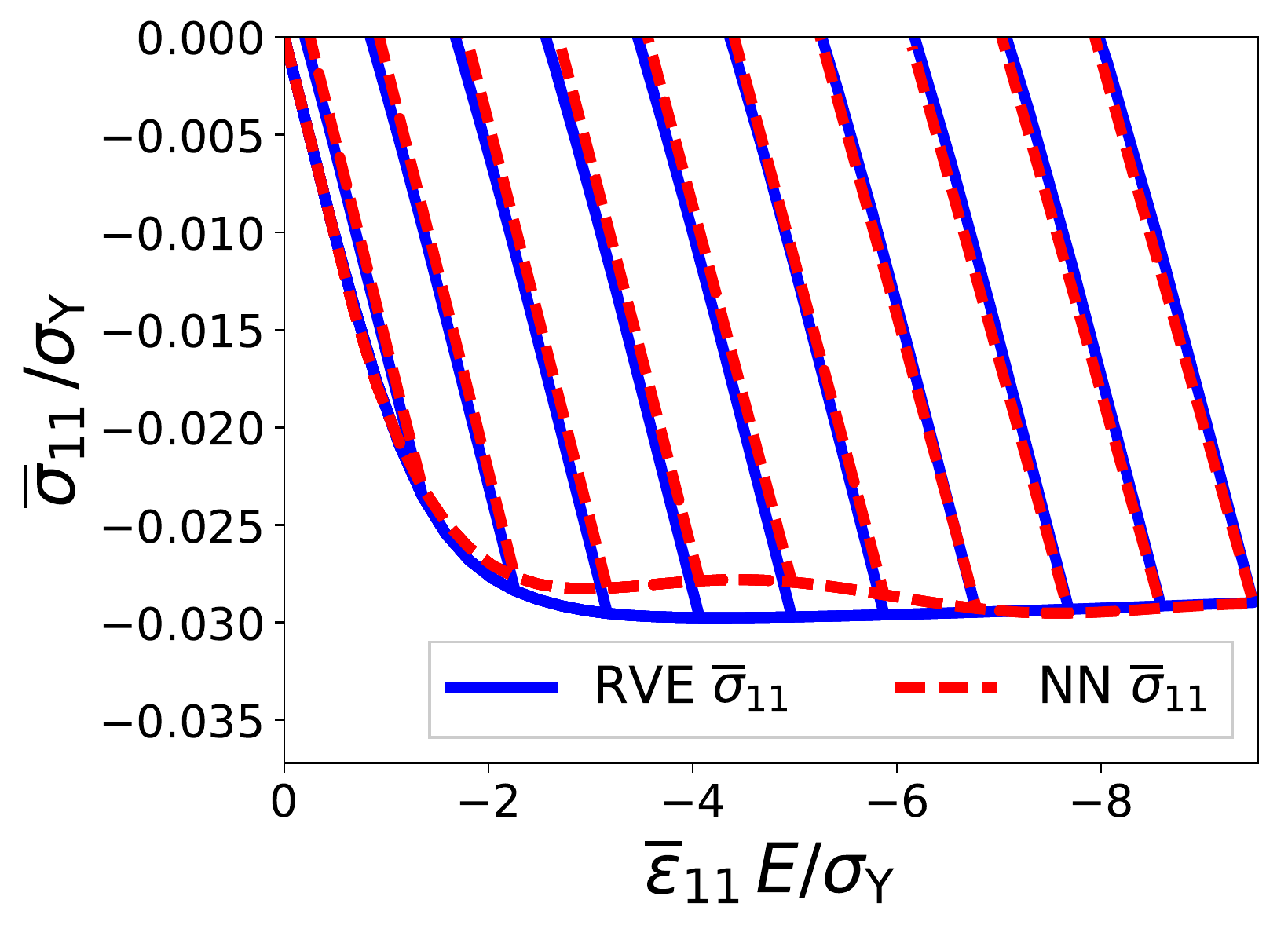}
	}
	\caption{Macroscopic stress-strain curves for different loading cases}
	\label{fig:MakroElasPlas_Schädigung_NN}
\end{figure}

Figure~\ref{fig:MakroElasPlas_Schädigung_NN} shows the response of the hybrid multi-scale neural network approach (\enquote{NN}) in comparison to the RVE simulations for three loading cases.
Obviously, both, the plastic behavior, manifested by remanent strains after unloading, as well as the damage, visible by an decreasing unloading stiffness, can be described adequately by HyMNNA.
Note that the stresses $\makro\sigma_{11}$ and $\makro\sigma_{22}$ are identical in the RVE simulations under bi-axial loading in \figurename~\ref{fig:MakroElasPlas_Schädigung_NN_Hydro} for symmetry reasons. In contrast, the HyMNNA yields a slight deviation between both values as can be seen in the right-hand side parts of Figs.~\ref{fig:MakroElasPlas_Schädigung_NN_Hydro}. 
The yield function in Eq.~\eqref{equ:MakElasPlas_MatMod_hybrid_Fliessfunk} is symmetric with respect to $\makro\sigma_{11}$ and $\makro\sigma_{22}$. Thus, the aforementioned deviation is attributed to the limited approximation accuracy of the yield direction in Eq.~\eqref{equ:MakElasPlas_MatMod_hybrid_Fliessrichtung}.

In contrast, the deviations under deviatoric loading (corresponding to pure shear \ang{45} to the coordinate axes) are attributed to the approximation accuracy of the yield function itself. In this context, it is recalled that the neural network was trained with respect to minimize the mean-square error of the yield function along all trained loading paths. However, \figurename~\ref{fig:MatMod_1RVE_hybrid_TeilEntlast_YieldSurfaces} had shown that the extension of the yield surface in hydrostatic direction {is larger} than in deviatoric directions (as it is well-known for foam materials). Thus, relative errors in deviatoric directions have less absolute impact on the mean-square error than relative errors in directions with high hydrostatic parts.
In future work, the accuracy in deviatoric directions could be improved by scaling down the hydrostatic contribution in the ansatz~\eqref{equ:MakElasPlas_MatMod_hybrid_Fliessfunk} for the yield function.
Anyway, the overall approximation quality of the HyMNN approach has a sufficient accuracy for most engineering applications, and their is still room for improvements of the accuracy.

\subsection{Plate with hole}

In the following, the application of the hybrid multi-scale neural network approach (HyMNNA) {is demonstrated for the heterogeneous deformation field of a rectangular plate with a hole. The plate is loaded uni-axially by uniform tractions $\makro{t}_2=0.88\,\makroFliessSpa$. This loading is slightly below the macroscopic yield stress $\makroFliessSpa=0.0228\mikro\FliessSpa$, so that plastic deformations will be present around the hole. Damage is not incorporated. 
For symmetry reasons, only a quarter model needs to be simulated as shown in \figurename~\ref{fig:Anw_Groessen_Verschiebung_Loch}.
\begin{figure}
	\centering
	\includegraphics{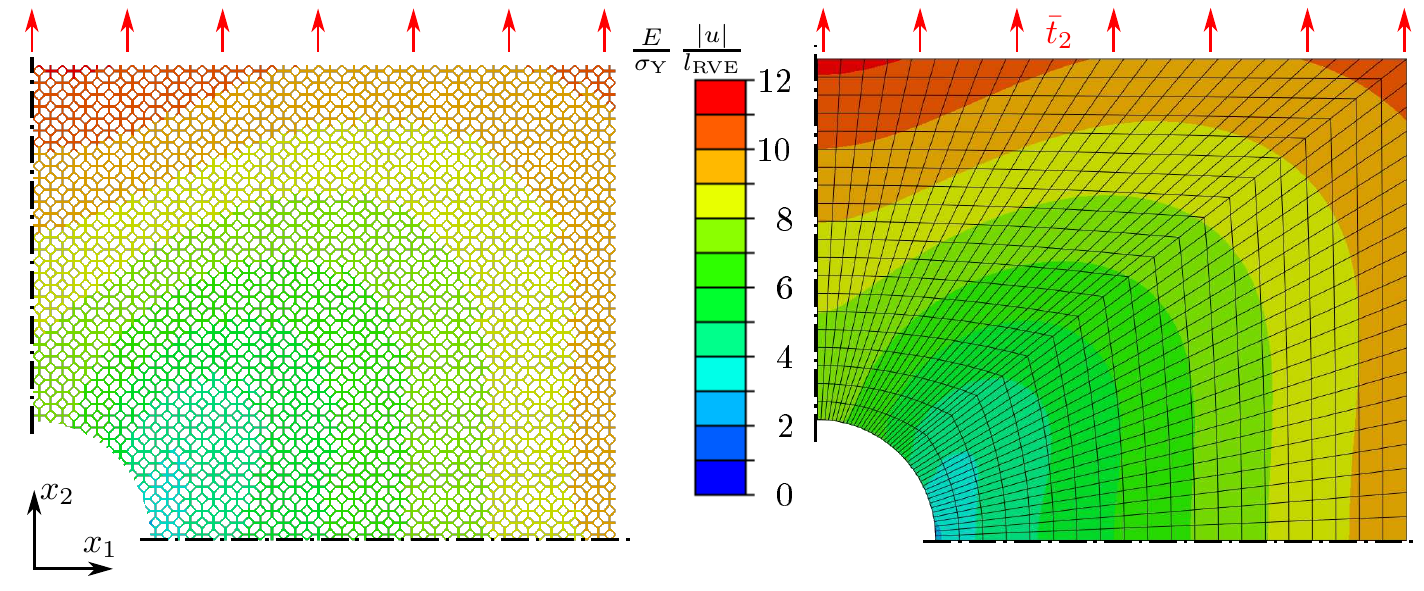}
	\caption{Displacement magnitude in DNS (left) and homogenized plate with hole (HyMNNA) (right)}
	\label{fig:Anw_Groessen_Verschiebung_Loch}
\end{figure}
For comparison, direct numerical simulations (DNS) with the discretely resolved micro-structure have been performed as shown at the left-hand side of \figurename~\ref{fig:Anw_Groessen_Verschiebung_Loch}. Within the DNS, the applied tractions are distributed statically equivalently to the rods of the foam.
The radius of the hole is chosen as $5\lRVE$. Such a ratio between radius and intrinsic material length $\lRVE$ is typically seen as lower limit for a homogenized theory to apply. Though, this choice leads to a DNS model with about 500 cells and about 2 million elements. In contrast, the simulation with HyMMNA requires only 900 elements and leads to a speed-up of factor 4000, even when the DNS is run on 15 CPU cores and HyMNNA on a single core. The speed-up would even been higher if a higher and more realistic scale separation ratio between hole radius and $\lRVE$ would be addressed. 
Figure~\ref{fig:Anw_Groessen_Verschiebung_Loch} shows a good agreement between both computed displacement fields.

The stress fields, which are compared in \figurename~\ref{fig:Anw_Groessen_Spannung_Loch}, are more sensitive to the constitutive relations.
\begin{figure}
	\centering
	\includegraphics{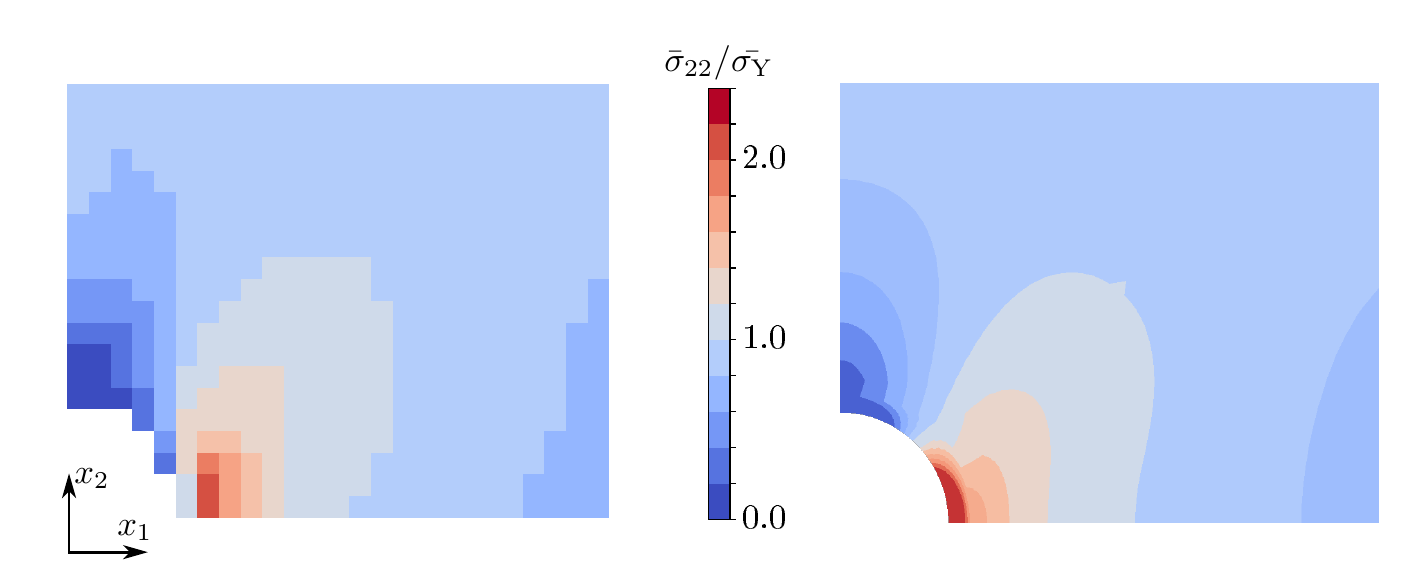}
	\caption{Stresses in DNS (left) and homogenized plate with hole (HyMNNA) (right)}
	\label{fig:Anw_Groessen_Spannung_Loch}
\end{figure}
The macroscopic stress fields for the DNS have been obtained by applying the averaging rule~\eqref{equ:Theorie_Homogen_makroDehnVolum} to the local fields within each cell of the DNS.
Further details on the extraction of the macroscopic values from DNS are provided in Appendix~\ref{app:DNS}.
The comparison shows that the stress fields can be captured quite well by the proposed HyMNN approach.
Only within the first volume element at the surface of the hole, the stresses deviate slightly. This behavior is attributed to the fact that the rods of the cell there are partly cut by the hole. Such surface layers cannot be described by the conventional Hill-Mandel homogenization theory,
but generalized continuum theories would be required to simulate such size effects, cf.~\citep{Tekoglu2011,Forest2005a}.
Anyway, the HyMNNA simulation provides good results at a tremendous speed-up compared to the DNS computation.}

\subsection{Foam filter}

\begin{figure}
	\centering
	\includegraphics[width=.85\columnwidth]{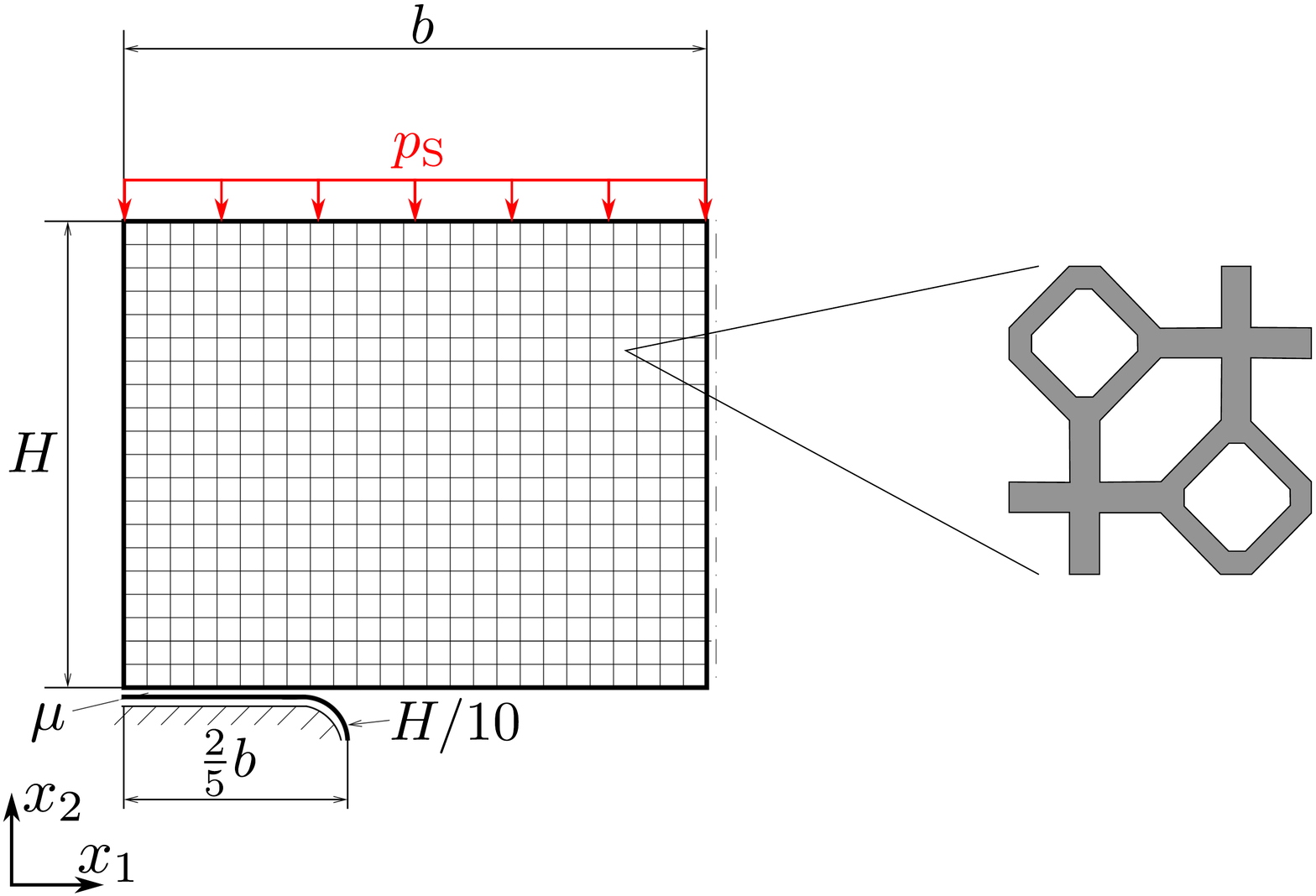}
	\caption{FEM Model of foam filter ($b=25\,\lRVE$, $\Hoehe=20\,\lRVE$) and structure of a single RVE}
	\label{fig:Anw_Modell}
\end{figure}

\begin{figure}
	\centering
	\includegraphics[trim=20 10 0 10,clip,height=68mm]{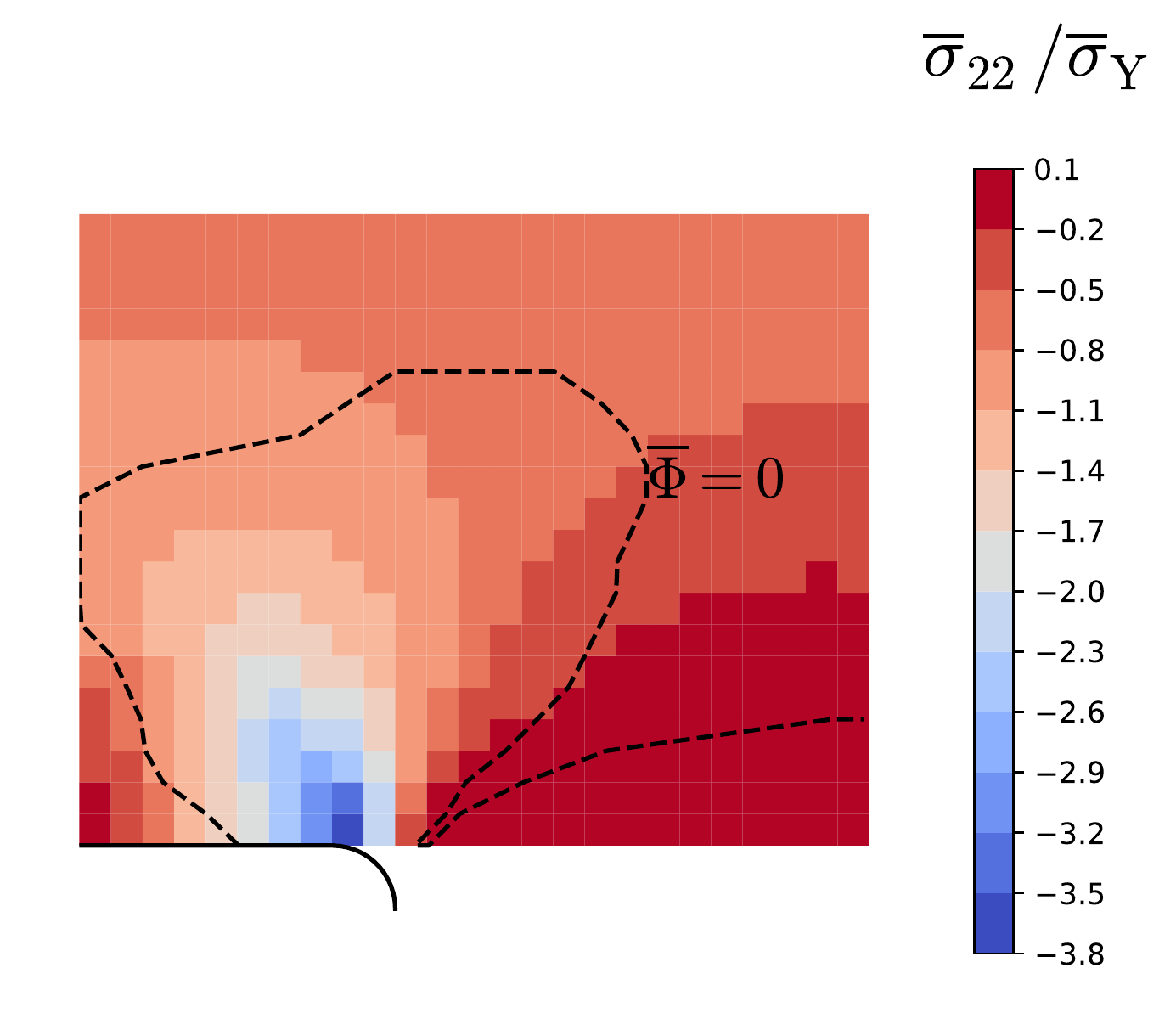}
	\hspace{5mm}
	\includegraphics[trim=20 10 90 10,clip,height=68mm]{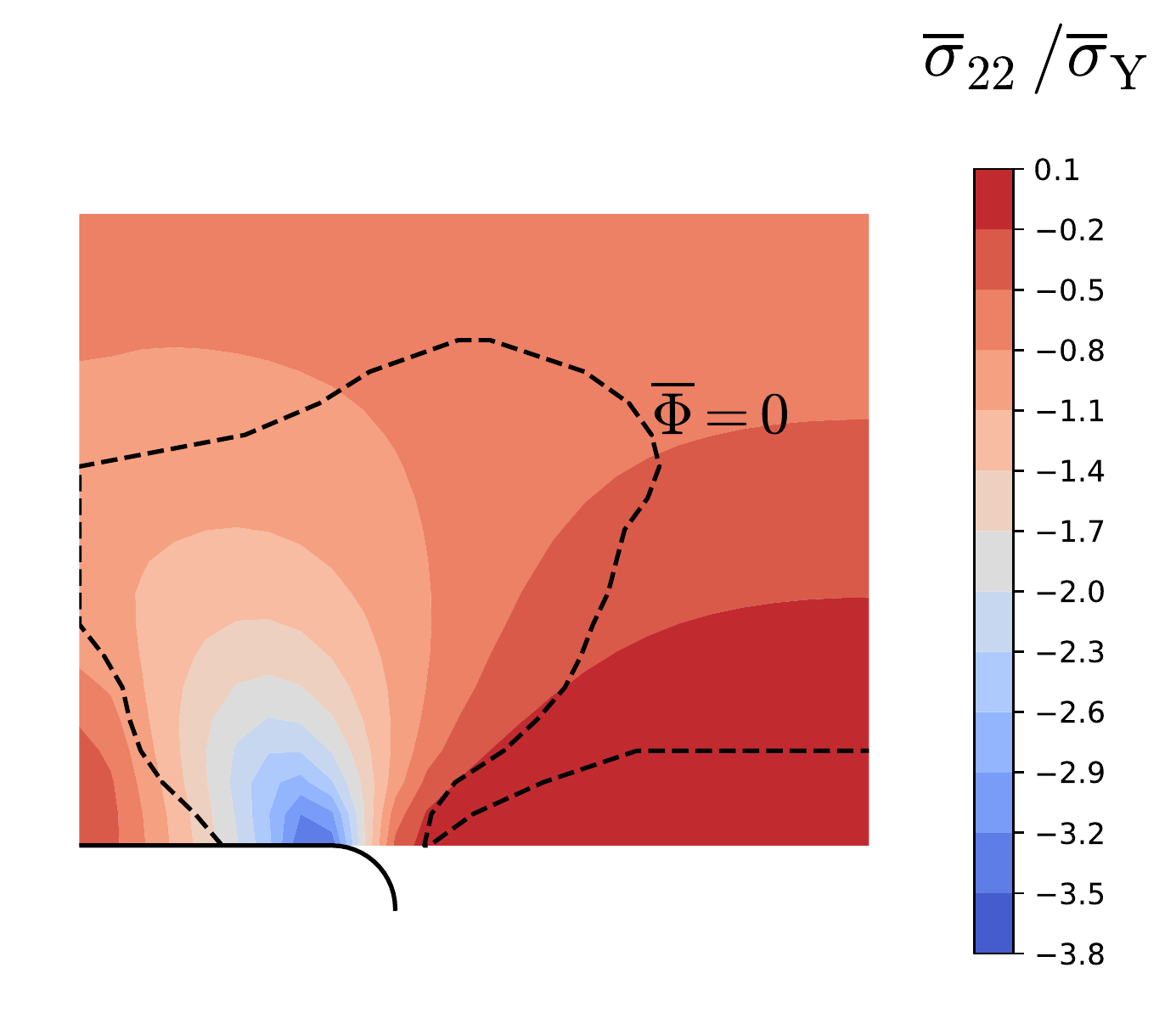}	
	\caption{Vertical macroscopic stress $\makro\sigma_{22}$ (normalized by macroscopic yield stress $\makro{\sigma}_\mathrm{Y} = 0.028\,\sigma_\mathrm{Y}$) and active plastic zone of DNS (left) and homogenized foam filter (HyMNNA) (right), both mapped to undeformed configuration}
	\label{fig:Anw_Groessen_Spannung}
\end{figure}

\begin{figure}
	\centering
	\includegraphics[trim=20 10 0 10,clip,height=68mm]{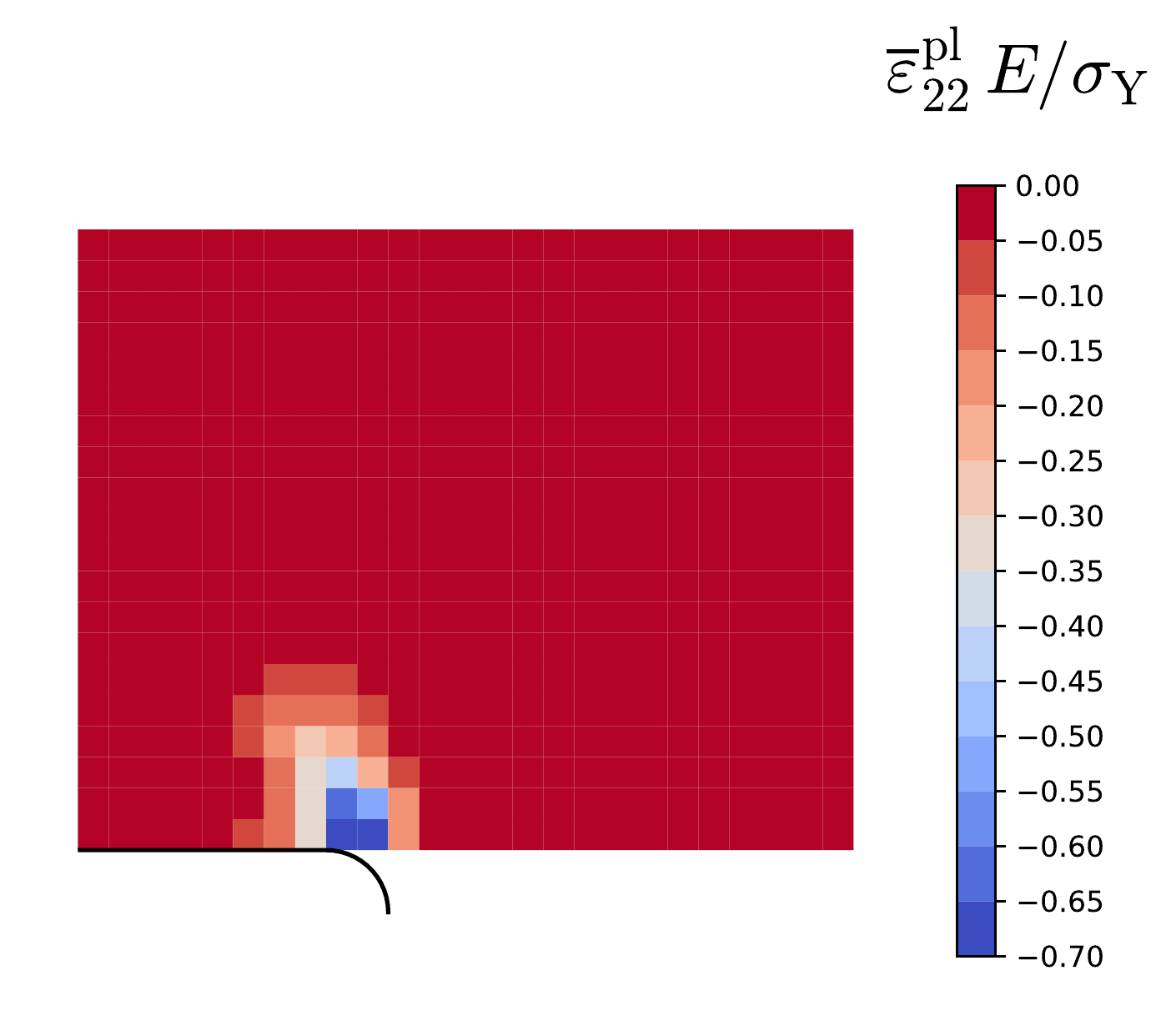}
	\hspace{5mm}
	\includegraphics[trim=20 10 103 10,clip,height=68mm]{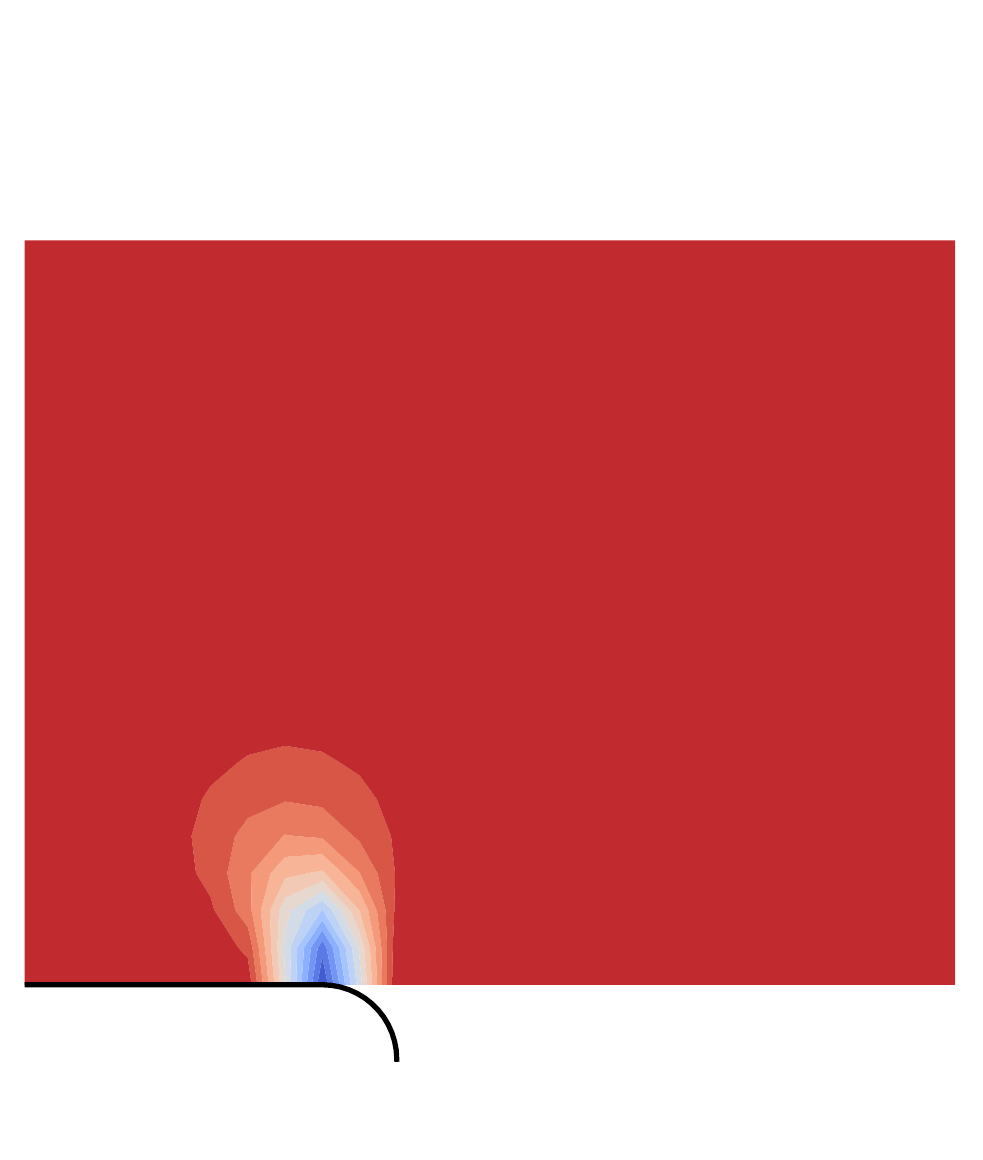}	
	\caption{Vertical macroscopic plastic strain $\makro\varepsilon_{22}^\mathrm{pl}$ of DNS (left) and homogenized foam filter (HyMNNA) (right), both mapped to undeformed configuration}
	\label{fig:Anw_Groessen_plastdehn}
\end{figure}

Now, the application of HyMNNA to the (idealized) problem of the elastic-plastic behavior of ceramic filters during filtration of steel melt shall be demonstrated.
\figurename~\ref{fig:Anw_Modell} shows the employed finite element mesh as well as loading and boundary conditions.
Due to mirror symmetry, only half of the model needs to be meshed. The bearing is assumed to be ideally rigid with a friction coefficient of $\mu=0.3$. The loading by the melt is modeled here by a constant pressure $p_\mathrm{S}=0.016\,\mikro\FliessSpa$. 
A direct numerical simulations (DNS) with the discretely resolved micro-structure is performed again for comparison.
Figures~\ref{fig:Anw_Groessen_Spannung} and \ref{fig:Anw_Groessen_plastdehn} show vertical components of stress and plastic strain, respectively, at the final step of DNS and homogenized solution by means of HyMNN approach. 
Furthermore, \figurename~\ref{fig:Anw_Groessen_Spannung} contains the active plastic zones, which initiate from the bearing.
In general, it can be said that the HyMNN approach reproduces the fields of stress and plastic strain quite well.
A comparison of left-hand sides and right-hand sides of both figures shows the accurate predictions of the HyMMN approach.

\subsection{Indentation}

\begin{figure}
	\centering
	\includegraphics{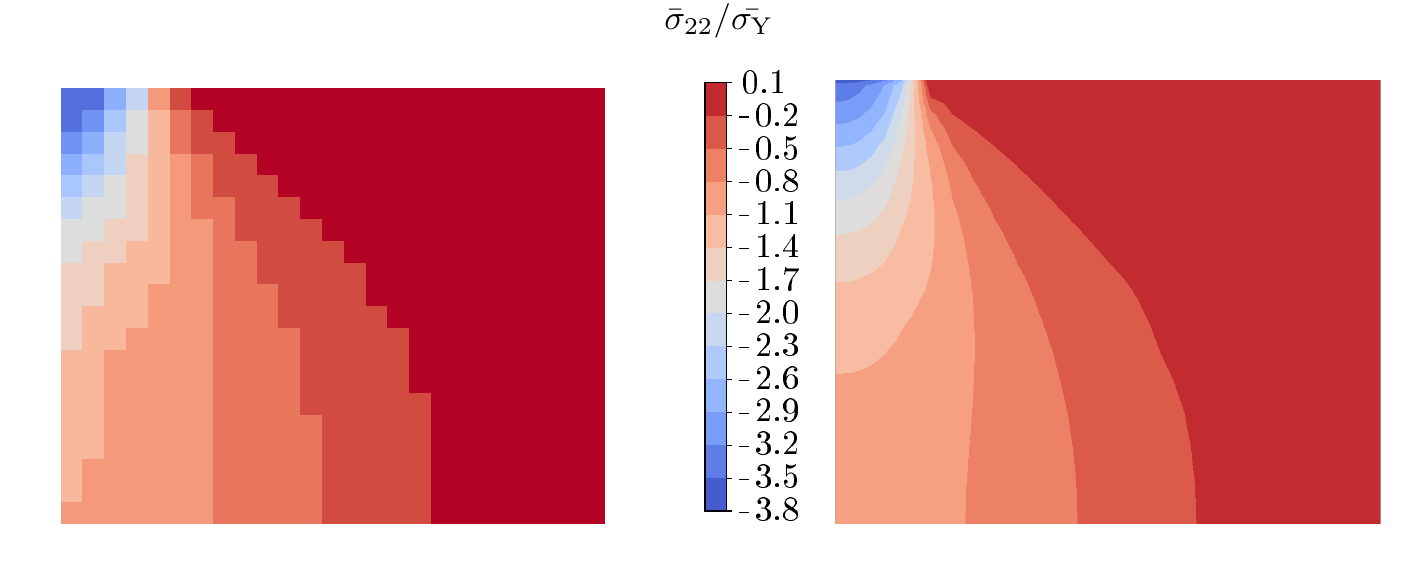}
	\caption{Vertical macroscopic stress $\makro\sigma_{22}$ (normalized by macroscopic yield stress $\makro{\sigma}_\mathrm{Y} = 0.028\,\sigma_\mathrm{Y}$) in DNS (left) and homogenized (HyMNNA) indentation problem (right) for $\rindent=20~\lRVE$ at $\uindent=\lRVE$, both mapped to undeformed configuration}
	\label{fig:Anw_Groessen_Spannung_Indenter}
\end{figure}

{As a last example, the indentation of an elastic-plastic foam is simulated. Geometry and boundary conditions are shown in \figurename~\ref{fig:Indenter}.
 The indented specimen has a size of 25~$\lRVE$ $\times$15~$\lRVE$. Two different values, $\rindent=10~\lRVE$ and $\rindent=20~\lRVE$, have been used for the radius of the circular indenter. Resulting load-deflection curves are plotted in \figurename~\ref{fig:Indenter_F_u}. The curves exhibit a good agreement between DNS and HyMMNA simulations. 
\begin{figure}
	\begin{minipage}[b]{0.46\textwidth}
		\centering
		\includegraphics{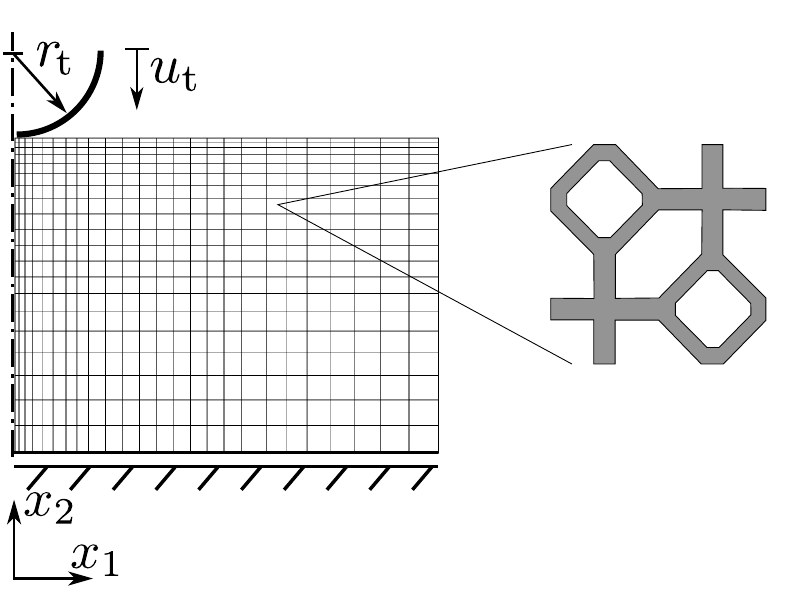}
		\caption{Indentation problem}
		\label{fig:Indenter}
	\end{minipage}
	\hfill
	\begin{minipage}[b]{0.46\textwidth}
		\centering
		\includegraphics{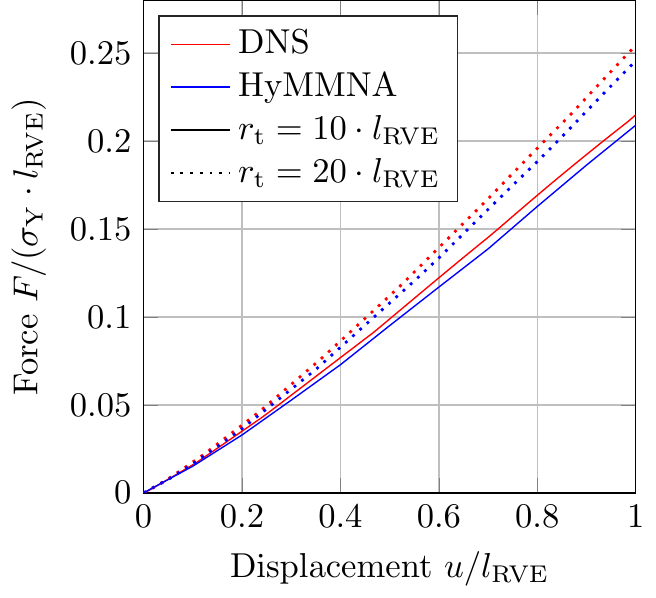}
		\caption{Load-deflection curves}
		\label{fig:Indenter_F_u}
	\end{minipage}
\end{figure}
Respective stress fields below the indenter are shown in \figurename~\ref{fig:Anw_Groessen_Spannung_Indenter}, demonstrating again a good agreement.}

\section{Summary and conclusions}
\label{sec:summary}

A new \enquote{hybrid} neural network macroscopic approach (HyMNNA) is presented to describe the irreversible material behavior at the macroscopic scale.
Known \emph{qualitative} features of the material behavior, like spatial symmetries, tension-compression symmetry or rate-independent behavior are taken into account by formulating the functional dependencies of the constitutive functions (state laws, potentials, yield functions, evolution equations), in the same way as it is done in conventional phenomenological material modeling.
In contrast to the latter, the constitutive functions are \emph{not} characterized by a certain number of constitutive parameters. Rather, the constitutive functions are represented by neural networks, whose response is generated from respective training data. 
The big advantage of this approach is that complex couplings within the functional dependencies, e.~g.\ of yield function and of evolution equations resulting for instance in distortional hardening effects and evolving anisotropy, do not need to be formulated in advance. Rather, such couplings are \enquote{learned} from the training data. 
Vice versa, this means that the set of training data needs to contain sufficient information about such couplings. Though, this is the case as well in conventional phenomenological modeling if respective coupling coefficients are to be determined.
The training is done \emph{offline}, i.~e.\ before the actual simulations.  During the simulations, only the relatively cheap evolution of the neural network response functions is required as \emph{online} computation. Thus, the performance of the hybrid multi-scale neural network approach and the effort for implementing it into finite element codes compare to conventional phenomenological constitutive models.

As an example, the hybrid neural network macroscopic approach was applied in the present proof-of-concept study to describe the constitutive behavior of anisotropic elastic-plastic foam structures with coupled anisotropic evolution of damage and non-associated flow rule. RVE simulations have been used as {synthetic} training data,
by which the elastic-plastic deformations of {a plate with a hole, of a foam filter and of an indentation test} could be simulated accurately and much more efficiently than a direct numerical simulation (DNS) or than a FE$^2$ simulation. 
Though, it should be possible as well to use the hybrid neural network macroscopic approach with experimental data for training as e.~g.\ in \citep{Ghaboussi1998ComputGeotech}.
In the present study a small deformation theory was used for simplicity, but the extension to large deformation is straightforward, if suitable measures of stress and strain are employed.

The dissipation rate was employed to judge the approximation quality of the neural network representation. Though, future studies need to address the estimation of errors in more detail. Furthermore, the implementation can be optimized with respect to hyperparameters of the neural network, like the number of neurons or of layers, respectively.

\section*{Acknowledgment}
The authors CS, MK and MA gratefully acknowledge the financial support by the Deutsche Forschungsgemeinschaft within the collaborative research center SFB 920 \enquote{Multi-functional filter for Metal Melt Filtration --- A Contribution towards Zero Defect Materials}. {The authors thank Alexander Malik for his help in performing the simulations in section~\ref{sec:application}.}

\bibliographystyle{elsarticle-harv}
\bibliography{HybridNNHomogenization}

\appendix

\section{Implementation}
\label{app:implementation}

The implemented stress update algorithm for the elastic-plastic model, Section~\ref{sec:elasticplasticmacro}, is outlined below in Algorithm~\ref{alg:hybrid2}. The regula falsi in the plastic branch requires two initial guesses for the increment of the equivalent plastic strain. The first one is chosen as $\Delta\makro\varepsilon_\mathrm{eq}^\mathrm{pl}=0$, corresponding to the elastic predictor. The norm of the strain increment $\Delta\makro\varepsilon_\mathrm{eq}^\mathrm{pl}=\Fnorm{\Delta\makro\varepsilon_{ij}}$ is used as second initial guess.
\begin{algorithm}[h!]
	\caption{Implemented stress update algorithm}
	\label{alg:hybrid2}
	\begin{algorithmic}[1]
		\LCOMMENT{Input of last values of stress and strain, strain increments and parameters}
		\STATE $\Moment{\makro\varepsilon_{ij}}{t}$, $\Moment{\makro\sigma_{ij}}{t}$, $\Delta\makro\varepsilon_{ij}$, $\Moment{\makro\varepsilon_\mathrm{eq}^\mathrm{pl}}{t}$, $\abs{\makro\varepsilon_{12}^\mathrm{pl}}_t$

		\LCOMMENT{Compute trial stress}
		\STATE $\Moment{\makro\sigma_{ij}}{t+\Delta t} = \Moment{\makro\sigma_{ij}}{t}+\makro{C}_{ijkl} \Delta\makro\varepsilon_{kl}$

		\LCOMMENT{Test whether plastic yielding occurs, compare Eq.~\eqref{equ:MakElasPlas_MatMod_hybrid_Fliessfunk}}
		\STATE $\makro{\Fliessfunk} = \sqrt{\Moment{\makro\sigma_{ij}}{t+\Delta t}\Moment{\makro\sigma_{ij}}{t+\Delta t}} - \NNfliess\left(\Moment{\makro{I}_1}{t+\Delta t},\Moment{\makro\varepsilon_\mathrm{eq}^\mathrm{pl}}{t},\abs{\makro\varepsilon_{12}^\mathrm{pl}}_t\right)$

		\IF{$\makro{\Fliessfunk}>0$}
		
		\LCOMMENT{\hspace{0.4cm}Compute increment of equivalent plastic strain~$\Delta\makro\varepsilon_\mathrm{eq}^\mathrm{pl}$ by means of \texttt{Regula falsi}-method as root of yield condition $\makro{\Fliessfunk}\stackrel{!}{=}0$ under consideration of updated values of stress \mbox{$\makro\sigma_{ij} = \Moment{\makro\sigma_{ij}}{t+\Delta t}-\Delta\makro\varepsilon_\mathrm{eq}^\mathrm{pl}\makro{C}_{ijkl}\makro{\breve{n}}_{kl}^\mathrm{pl}$} and flow direction \mbox{$\dilatflowangle =  \NNdirec\left(\makro{I}_1,\makro\varepsilon_\mathrm{eq}^\mathrm{pl},\abs{\makro\varepsilon_{12}^\mathrm{pl}}\right)$} according to Eq.~\eqref
			{equ:MakElasPlas_MatMod_hybrid_Fliessrichtung}}
		\STATE $\Delta\makro\varepsilon_\mathrm{eq}^\mathrm{pl} = \texttt{RegulaFalsi}\left( \Moment{\makro\sigma_{ij}}{t+\Delta t},\Moment{\dilatflowangle}{t+\Delta t},\Moment{\makro\varepsilon_\mathrm{eq}^\mathrm{pl}}{t+\Delta t},\abs{\makro\varepsilon_{12}^\mathrm{pl}}_{t+\Delta t} \right)$
		
		\LCOMMENT{\hspace{0.4cm}Update within each iteration}
		\STATE $\Moment{\makro\varepsilon_\mathrm{eq}^\mathrm{pl}}{t+\Delta t} = \Moment{\makro\varepsilon_\mathrm{eq}^\mathrm{pl}}{t} + \Delta\makro\varepsilon_\mathrm{eq}^\mathrm{pl}$
		\STATE $\abs{\makro\varepsilon_{12}^\mathrm{pl}}_{t+\Delta t}=\abs{\makro\varepsilon_{12}^\mathrm{pl}}_t + \Delta\makro\varepsilon_\mathrm{eq}^\mathrm{pl}\abs{\makro{\breve{n}}_{12}^\mathrm{pl}}$
		\STATE $\Moment{\makro\sigma_{ij}}{t+\Delta t} = \Moment{\makro\sigma_{ij}}{t+\Delta t}-\Delta\makro\varepsilon_\mathrm{eq}^\mathrm{pl}\makro{C}_{ijkl}\makro{\breve{n}}_{kl}^\mathrm{pl}$
		
		\ELSE
		
		\STATE $\Moment{\makro\varepsilon_\mathrm{eq}^\mathrm{pl}}{t+\Delta t} = \Moment{\makro\varepsilon_\mathrm{eq}^\mathrm{pl}}{t}$; \quad $\abs{\makro\varepsilon_{12}^\mathrm{pl}}_{t+\Delta t}=\abs{\makro\varepsilon_{12}^\mathrm{pl}}_t$

		\ENDIF
		
		\LCOMMENT{Compute dissipation according to Eq.~\eqref
			{equ:MakElasPlas_MatMod_AendDissi}}
		\STATE $\Delta{\makro\Dissi} = \Delta\makro\varepsilon_\mathrm{eq}^\mathrm{pl} \Moment{\makro\sigma_{ij}}{t+\Delta t}\Moment{\makro{\breve{n}}_{ij}^\mathrm{pl}}{t+\Delta t}$
		\label{alg:hybrid2_AendDissi}

		\LCOMMENT{Output: Updated stress, plastic strains and dissipation}
		\STATE $\Moment{\makro\sigma_{ij}}{t+\Delta t}$, $\Moment{\makro\varepsilon_\mathrm{eq}^\mathrm{pl}}{t+\Delta t}$, $\abs{\makro\varepsilon_{12}^\mathrm{pl}}_{t+\Delta t}$, $\Delta{\makro\Dissi}$
	\end{algorithmic}
\end{algorithm} 

\section{Extraction of macroscopic fields from DNS}
\label{app:DNS}

The macroscopic stress components $\makro\sigma_{ij}$ are computed from Eq.~\eqref{equ:Theorie_Homogen_makroDehnVolum}$_1$ for each RVE (as shown in the inset of \figurename~\ref{fig:Theorie_Homogen_MikroMakro}) in the DNS. Thereby, a vanishing microscopic stress $\mikro\sigma_{ij}=0$ can be assigned unambiguously to the pores. These values of $\makro\sigma_{ij}$ are assigned to the center of each RVE. 

The extraction of the strain components $\makro\varepsilon_{ij}$ is a little bit more involved since \emph{no unique} strain field $\mikro\varepsilon_{ij}$ can be assigned to the pore, so that the average operator in Eq.~\eqref{equ:Theorie_Homogen_makroDehnVolum}$_2$ cannot strictly be used to extract $\makro\varepsilon_{ij}$. 
Sometimes, this problem is addressed by assigning an elastic material with low Young's modulus to the pore. 
Though, this method requires additionally to assign a value for the Poisson ratio $\nu$ to the pore.
This method is not used here since then the microscopic strain field $\mikro\varepsilon_{ij}$ and thus the extracted $\makro\varepsilon_{ij}$ would depend in general on this arbitrarily chosen Poisson ratio $\nu$ of the pore.

\begin{figure}
	\centering
	\includegraphics[width=0.5\textwidth]{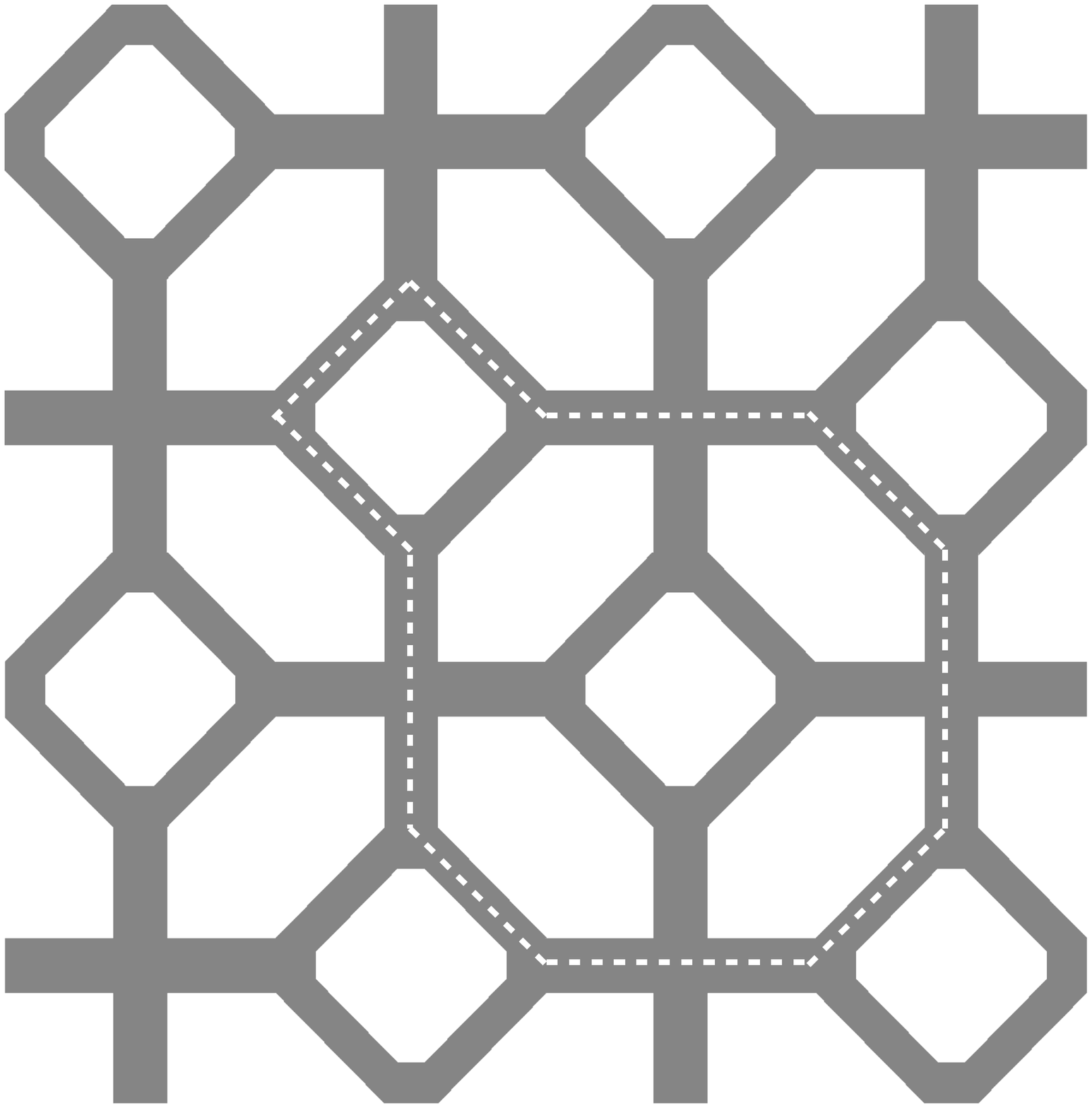}
	\caption{Path $\partial V_\mathrm{RVE}$ for extraction of macroscopic strains $\makro\varepsilon_{ij}$ from DNS}
	\label{fig:RVE_path_through_matrix}
\end{figure}

Assuming that at least a sufficiently smooth displacement field can be defined (not necessarily uniquely) in the pores, the volume integral in Eq.~\eqref{equ:Theorie_Homogen_makroDehnVolum}$_2$ can be transformed to a surface integral
\begin{equation}
	\makro\varepsilon_{ij} = \frac{1}{2\,V_\mathrm{RVE}} \underset{\partial V_\mathrm{RVE}}{\oint} \left(u_i n_j + n_i u_j\right) \mathrm{d}A
	\label{equ:Theorie_Homogen_makroDehn}
\end{equation}
over a closed surface $\partial V_\mathrm{RVE}$ of the RVE. The choice of the RVE $V_\mathrm{RVE}$ is not unique, but any periodically continuable part of the microstructure can be chosen as $V_\mathrm{RVE}$. The surface integral in Eq.~\eqref{equ:Theorie_Homogen_makroDehn} requires to have a unique displacement field $u_i$ defined at $\partial V_\mathrm{RVE}$. Consequently, unique values of macroscopic strain $\makro\varepsilon_{ij}$ can be extracted from the present DNS (or any other closed-cell foam), if $V_\mathrm{RVE}$ is chosen such that its boundary $\partial V_\mathrm{RVE}$ goes solely through the matrix material, but not through a pore as shown in \figurename~\ref{fig:RVE_path_through_matrix}.
Having extracted the total strain, its plastic part can be computed by Eq.~\eqref{equ:statelawmacroinvertedplastic}.

\end{document}